\documentclass[journal]{IEEEtran}

\def\mode{0}  

\def\redactionmode{1} 

\def\alttextmode{1} 
\usepackage{adjustbox}              
\usepackage{amsmath}
\usepackage{cite}					
\usepackage{cleveref}
\usepackage{comment}  				

\usepackage{enumitem}				
\usepackage{graphicx}
	\graphicspath{{./figures/}}		
\usepackage{multirow}				
\usepackage{multicol}
\usepackage{tabularx} 				
\usepackage{threeparttable}
\usepackage{soul} 					
\usepackage{xargs}    				
\usepackage[svgnames]{xcolor}			

\newcommand{\red}{}
\newcommand{\green}{}
\newcommand{\blue}{}

\newcounter{documentmode}
\includecomment{conf}  	
\newcommand{\setdocmode}{%
  \ifcase\number\value{documentmode}
	\renewcommand{\red}[1]{\color{red}##1\color{black}{}}
	\renewcommand{\green}[1]{\color{Green}##1\color{black}{}}
	\renewcommand{\blue}[1]{\color{blue}##1\color{black}{}}
  \or
    \usepackage{draftwatermark}
	\SetWatermarkScale{1.35}
	\SetWatermarkAngle{56}
	\SetWatermarkLightness{0.9}
	\SetWatermarkHorCenter{0.9\textwidth}
	\SetWatermarkVerCenter{0.7\textheight}
  \or
	\usepackage{draftwatermark}
	\SetWatermarkText{CONFIDENTIAL}
	\SetWatermarkScale{0.67}
	\SetWatermarkAngle{56}
	\SetWatermarkLightness{0.9}
	\SetWatermarkHorCenter{0.6\textwidth}
	\SetWatermarkVerCenter{0.6\textheight}
  \or
	\usepackage{draftwatermark}
	\SetWatermarkText{Pre-Publication}
	\SetWatermarkScale{2.67}
	\SetWatermarkAngle{56}
	\SetWatermarkLightness{0.9}
	\SetWatermarkHorCenter{0.6\textwidth}
	\SetWatermarkVerCenter{0.6\textheight}
  \else
    \excludecomment{conf}
  \fi
}

\if\mode0
\usepackage[prependcaption, textsize=footnotesize]{todonotes}
\else
\usepackage[disable, prependcaption, textsize=footnotesize]{todonotes}

\fi

\setdocmode

\newcounter{redmode}
\includecomment{2Detailed}			
\includecomment{Uncritical}			
\includecomment{2SaveSpace}			
\includecomment{IfMustDelete}			
\newcommand{\redact}{%
  \ifcase\number\value{redmode}
    \includecomment{2Detailed}
    \includecomment{Uncritical}
    \includecomment{2SaveSpace}
    \includecomment{IfMustDelete}
  \or
    \excludecomment{2Detailed}
    \includecomment{Uncritical}
    \includecomment{2SaveSpace}
    \includecomment{IfMustDelete}
  \or
    \excludecomment{2Detailed}
    \excludecomment{Uncritical}
    \includecomment{2SaveSpace}
    \includecomment{IfMustDelete}
  \or
    \excludecomment{2Detailed}
    \excludecomment{Uncritical}
    \excludecomment{2SaveSpace}
    \includecomment{IfMustDelete}
  \or
    \excludecomment{2Detailed}
    \excludecomment{Uncritical}
    \excludecomment{2SaveSpace}
    \excludecomment{IfMustDelete}
  \fi
}
\redact

\newcounter{alttxtmode}
\includecomment{TextA}			
\includecomment{TextB}			
\newcommand{\alttext}{%
  \ifcase\number\value{alttxtmode}
    \includecomment{TextA}
    \includecomment{TextB}
  \or
    \excludecomment{TextA}
    \includecomment{TextB}
  \or
    \includecomment{TextA}
    \excludecomment{TextB}
  \or
    \excludecomment{TextA}
    \excludecomment{TextB}
  \fi
}
\alttext

\def\Ron{\ensuremath{R_\mathrm{on}}}
\def\Roff{\ensuremath{R_\mathrm{off}}}

\usetikzlibrary{shadows.blur}
\definecolor{turquoise}{RGB}{29,147,153}
\definecolor{dark blue}{RGB}{29,147,153}
\definecolor{petrol}{RGB}{14,97,128}
\definecolor{light blue}{RGB}{24,169,224}
\definecolor{darkgray}{RGB}{68,68,68}
\definecolor{light red}{RGB}{255,70,85}
\def\angle{90}
\def\radius{2}
\newcommand\squaretwo[1]{%
  \tikz\filldraw[color=#1] (0,0) rectangle (1.5ex,1.5ex);%
}
\newcount\ind \ind=-1

\usepackage[utf8]{inputenc}
\usepackage[europeanresistors,RPvoltages]{circuitikz}
\usetikzlibrary{positioning}

\usepackage{subfigure}
\usepackage{nccmath}
\usepackage{cuted}

\usepackage[acronym]{glossaries} 

\newacronym{am}{AM}{Amplitude Modulation}
\newacronym{asd}{ASD}{Autism Spectrum Disorder}
\newacronym{at}{AT}{Adaptive Tracking}
\newacronym{bw}{BW}{Baseline Wanderer}
\newacronym{bpm}{BPM}{Breath per Minute}
\newacronym{bvp}{BVP}{Blood Volume Pulse}
\newacronym{cemd}{CEMD}{Complex Empirical Mode Decomposition}
\newacronym{co}{CO}{Count Origin}
\newacronym{cosf}{TDCO}{Time Domain - Count Origin}
\newacronym{costf}{COSTF}{Count Origin - Smart and Time Fusion}
\newacronym{cwt}{CWT}{Continuous Wavelet Transform}
\newacronym{der}{DER}{Detection Error Rate}
\newacronym{df}{DF}{Dominant Frequency}
\newacronym{dfsf}{SFFDP}{Smart Fusion of Frequency Domain Peak}
\newacronym{dwt}{DWT}{Discrete Wavelet Transform}
\newacronym{ecg}{ECG}{Electrocardiography}
\newacronym{skin_eda}{EDA}{Electrodermal activity} 
\newacronym{eeg}{EEG}{Electroencephalogram}
\newacronym{emd}{EMD}{Empirical Mode Decomposition}
\newacronym{emg}{EMG}{Electromyography}
\newacronym{eog}{EOG}{Electrooculogram}
\newacronym{ews}{EWS}{Early Warning Score}
\newacronym{fft}{FFT}{Fast Fourier Transformation}
\newacronym{fm}{FM} {Frequency Modulation}
\newacronym{gsr}{GSR}{Galvanic Skin Response}
\newacronym{hf}{HF}{High Frequency}
\newacronym{hr}{HR}{Heart Rate}
\newacronym{hrv}{HRV}{Heart rate variability}
\newacronym{ibi}{IBI}{Inter beat interval}
\newacronym{ica}{ICA}{Independent Component Analysis}
\newacronym{iir}{IIR}{Infinite Impulse Response}
\newacronym{imf}{IMF}{Intrinsic Mode Function}
\newacronym{knn}{KNN}{k-nearest neighbour}
\newacronym{lda}{LDA}{Linear Discriminant Analysis}
\newacronym{led}{LED}{light emitting diode}
\newacronym{lms}{LMS}{Least Mean Square}
\newacronym{lf}{LF}{low frequency}
\newacronym{ma}{MA}{Movement Artifact}
\newacronym{mad}{MAD}{Mean Absolute Deviation} 
\newacronym{mar}{MAR}{Motion Artifact Removal} 
\newacronym{mimic}{MIMIC}{Multi-parameter Intelligent Monitoring for Intensive Care}
\newacronym{nlms}{NLMS}{Normalized Least Mean Square}
\newacronym{n2n}{N2N}{normal-to-normal}
\newacronym{pd}{PD}{Peak Detection}
\newacronym{pdsf}{TDPD}{Time Domain Peak Detection}
\newacronym{ppe}{PPE}{Peak to Peak Error}
\newacronym{ppg}{PPG}{Photoplethysmogram}
\newacronym{ptt}{PTT}{Pulse transit time}
\newacronym{rls}{RLS}{ Recursive Least Square}
\newacronym{rms}{RMS}{ Root Mean Square}
\newacronym{rmse}{RMSE}{Root Mean Square Error}
\newacronym{rr}{RR}{Respiratory Rate}
\newacronym{rrsv}{RRSV}{Repeated Random Sub-Sampling Validation}
\newacronym{sam}{SAM}{Self-Assessment Manikin}
\newacronym{saews}{SA-EWS}{Self-Aware Early Warning Score}
\newacronym{sart}{SART}{sustained attention to response test}
\newacronym{sfu}{SFU}{Smart Fusion}
\newacronym{scl}{SCL}{Skin Conductance Level}
\newacronym{scr}{SCR}{Skin Conductance Response}
\newacronym{sigvm}{SigVM}{Signal Vector Magnitude}
\newacronym{skt}{SKT}{Skin Temperature}
\newacronym{std}{STD}{Standard Deviation }
\newacronym{svd}{SVD}{Singular Value Decomposition }
\newacronym{tfu}{TFU}{Temporal Fusion } 
\newacronym{vfcdm}{VFCDM} {Variable Frequency Complex Demodulation}
\newacronym{whs}{WHS}{Wearable Health-care Systems}

\newacronym{ai}{AI}{Artificial Intelligence}
\newacronym{ann}{ANN}{Artificial Neural Network}
\newacronym{bpn}{BPN}{Back-Propagation Neural Network}
\newacronym{bpnn}{BPNN}{back-propagation neural network}
\newacronym{cart}{CART}{Classification And Regression Tree}
\newacronym{cnn}{CNN}{Convolutional Neural Network}
\newacronym{flop}{FLOP}{Floating Point Operation}
\newacronym{icnn}{ICNN}{Iterative Convolutional Neural Network}
\newacronym{ldf}{LDF}{Linear Discriminant Function}
\newacronym{mcs}{MCS}{Multiple Classifier System}
\newacronym{ml}{ML}{Machine Learning}
\newacronym{nn}{NN}{Neural Network}
\newacronym{nb}{NB}{Naive Bayesian}
\newacronym{rsvm}{RSVM}{Reputation-driven Support Vector Machine }
\newacronym{svm}{SVM}{Support Vector Machine}
\newacronym{ucnn}{$\mu$CNN}{Micro CNN}

\newacronym{brs}{BRS}{Bipolar Resistive Switch-based logic}
\newacronym{cnf}{CNF}{Conjunctive Normal Form}
\newacronym{crs}{CRS}{Complementary Resistive Switch-based logic}
\newacronym{dnf}{DNF}{Disjunctive Normal Form}
\newacronym{fpm}{FPM}{Forward Polarized Memristor}
\newacronym{hfo}{$HfO_x$}{Hafnium Oxide}
\newacronym{hrs}{HRS}{High Resistance State}
\newacronym{imc}{IMC}{In-Memory Computation}
\newacronym{imply}{IMPLY}{Material Implication}
\newacronym{imp}{IMP}{In-Memory Processing}
\newacronym{lim}{LIM}{Logic in Memory}
\newacronym{lrs}{LRS}{Low Resistance State}
\newacronym{ltg}{LTG}{Logic Threshold Gate}
\newacronym{magic}{MAGIC}{Memristor-Aided Logic}
\newacronym{mecoins}{Me-Coin}{Memristor-based Computation In-memory}
\newacronym{pcm}{PCM}{Phase Change Memory}
\newacronym{pim}{PIM}{Processing in Memory}
\newacronym{reram}{ReRAM}{Resistive Random Access Memory}
\newacronym{rpm}{RPM}{Reversely Polarized Memristor}
\newacronym{stt}{STT}{Spin Transfer Torque}
\newacronym{tao}{$TaO_x$}{Tantalum Oxide}
\newacronym{tio}{$TiO_2$}{Titanium dioxide}
\newacronym{vteam}{VTEAM}{Voltage-controlled ThrEshold Adaptive Memristor}
\newacronym{team}{TEAM}{ThrEshold Adaptive Memristor}

\newacronym{aco}{ACO}{Autonomous Cooperating Object}
\newacronym{afdd}{AFDD}{Automated Fault Detection and Diagnostic}
\newacronym{ca}{CA}{Continuous Average}
\newacronym{cah}{CAH}{Context-Aware Health Monitoring}
\newacronym{cam}{CCAM}{Confidence-based Context-Aware condition Monitoring}
\newacronym{csa}{CSA}{Computational Self-Awareness}
\newacronym[plural=DABs,longplural={Discrete Average Blocks}]{dab}{DAB}{Discrete Average Block}
\newacronym[plural=KPNs,longplural={Kahn Process Networks}]{kpn}{KPN}{Kahn Process Networks} 
\newacronym{mape-k}{MAPE-K}{Monitor-Analyze-Plan-Execute over a shared Knowledge}
\newacronym[plural=MoCs,longplural={Models of Computation}]{moc}{MoC}{Model of Computation}
\newacronym{oda}{ODA}{Observe-Decide-Act}
\newacronym{pca}{PCA}{Principal Component Analysis}
\newacronym{rosa}{RoSA}{Research on Self-Awareness}
\newacronym{sa}{SA}{Self-Aware}
\newacronym{saness}{SA}{Self-Awareness}
\newacronym{sahm}{SAHM}{Self-Aware Health Monitoring}
\newacronym{samba}{SAMBA}{Self-Aware health Monitoring and Bio-inspired coordination for distributed Automation systems}
\newacronym{selphys}{SelPhyS}{Self-aware cyber-Physical System}
\newacronym{sh}{SH}{State Handler}
\newacronym{som}{SOM}{Self-Organizing Map}

\newacronym{c-s}{CAS}{Compare-and-Swap}
\newacronym{cps}{CPS}{Cyber-Physical System}
\newacronym{cpps}{CPPS}{Cyber-Physical Production System}
\newacronym{dsr}{DSR}{Down-Sampling Rate}
\newacronym{dum}{DuM}{Device under Monitoring}
\newacronym{es}{ES}{Embedded System}
\newacronym{LD}{LD}{low-discrepancy}
\newacronym{mes}{MES}{Manufacturing Execution System}
\newacronym{sc}{SC}{Stochastic Computing}
\newacronym{sos}{SoS}{System of Systems}
\newacronym{suo}{SuO}{System under Observation}

\newacronym{abi}{ABI}{Application Binary Interface}
\newacronym{adc}{ADC}{Analog-to-Digital Converter}
\newacronym{aes}{AES}{Advanced Encryption Standard}
\newacronym{alu}{ALU}{Arithmetic Logic Unit}
\newacronym{api}{API}{Application Programming Interface}
\newacronym{asic}{ASIC}{Application Specific Integrated Circuit}
\newacronym{asoc}{ASOC}{Autonomic System-on-Chip platform}
\newacronym{axi}{AXI}{Advanced eXtensible Interface Bus}
\newacronym{bram}{BRAM}{Block Random Access Memory}
\newacronym{cdt}{CDT}{C/C++ Development Tooling}
\newacronym{clb}{CLB}{Configuarable Logic Block}
\newacronym{cmos}{CMOS}{Complementary Metal-Oxide Semiconductor}
\newacronym{cp}{CP}{Clock Pulse}
\newacronym{cpi}{CPI}{Cycles Per Instruction}
\newacronym{cpu}{CPU}{Central Processing Unit} 
\newacronym{cpsoc}{CPSoC}{Cyber-Physical System-on-Chip}
\newacronym{cu}{CU}{Compute Unit}
\newacronym{cuda}{CUDA}{Compute Unified Device Architecture}
\newacronym{dac}{DAC}{Digital to Analog Converter}
\newacronym{ddr3}{DDR3}{Double Data Rate}
\newacronym{dff}{DFF}{Data Flip-Flop}
\newacronym{dll}{DLL}{Delay Locked Loop}
\newacronym{dmr}{DMR}{Dual Modular Redundancy}
\newacronym{dram}{DRAM}{Dynamic Random Access Memory}
\newacronym{dsd}{DSD}{Digital Synchronous Detection}
\newacronym{dsp}{DSP}{Digital Signal Processor}
\newacronym{dt}{DigiTime}{}
\newacronym{dvfs}{DVFS}{Dynamic Voltage and Frequency Scaling}
\newacronym{eda}{EDA}{Electronic Design Automation}
\newacronym{fdc}{FDC}{Frequency-to-Digital Converter}
\newacronym{fifo}{FIFO}{First In First Out}
\newacronym{fpga}{FPGA}{Field Programmable Gate Array}
\newacronym{gds}{GDS}{Global Data Share}
\newacronym{gnulgpl}{GNU LGPL}{GNU Lesser General Public Licence} 
\newacronym{gpgpu}{GPGPU}{General Purpose Graphics Processing Unit}
\newacronym{gpr}{GPR}{General Purpose Register}
\newacronym{gpu}{GPU}{Graphics Processing Unit}
\newacronym{gro}{GRO}{Gated Ring Oscillator}
\newacronym{io}{IO}{Input-Output}
\newacronym{hamsoc}{HAMSoC}{Hierarchical Agent Monitoring System-on-Chip}
\newacronym{hdl}{HDL}{Hardware Description Language}
\newacronym{hmp}{HMP}{Heterogeneous Multi-Processor}
\newacronym{ic}{IC}{Integrated Circuit}
\newacronym{icap}{ICAP}{Internal Configuration Access Port}
\newacronym[longplural={Intellectual Properties}]{ip}{IP}{Intellectual Property}
\newacronym{isa}{ISA}{Instruction Set Architecture}
\newacronym{lds}{LDS}{Local Data Share}
\newacronym{lru}{LRU}{Least Recently Used}
\newacronym{lsb}{LSB}{Least-Significant Bit}
\newacronym{lsu}{LSU}{Load Store Unit}
\newacronym{lut}{LUT}{Look Up Table}
\newacronym{mash}{MASH}{Multi-Stage Noise-Shaping}
\newacronym{mems}{MEMS}{Micro-Electro-Mechanical Systems}
\newacronym{miaow}{MIAOW}{Many-core Integrated Accelerator Of deepwater/Wisconsin}
\newacronym{mosfet}{MOSFET}{Metal Oxide Semiconductor Field Effect Transistor}
\newacronym{mpsoc}{MPSoC}{Multi-Processor System-on-Chip}
\newacronym{mshr}{MSHR}{Miss Status Holding/Handling Register}
\newacronym{noc}{NoC}{Network-on-Chip}
\newacronym{opencl}{OpenCL}{Open Computing Language}
\newacronym{ocn}{OCN}{On-Chip Network}
\newacronym{pcb}{PCB}{Printed Circuit Board}
\newacronym{pcie}{PCIe}{Peripheral Component Interconnect Express}
\newacronym{pl}{PL}{Programmable Logic}
\newacronym{pli}{PLI}{Verilog Programming Language Interface}
\newacronym{pll}{PLL}{Phase-Locked Loop}
\newacronym{ps}{PS}{Processing System}
\newacronym{pv}{PV}{Process Variation}
\newacronym{qoe}{QoE}{Quality of Experience}
\newacronym{qos}{QoS}{Quality of Service}
\newacronym{ram}{RAM}{Random Access Memory} 
\newacronym{risc}{RISC}{Reduced Instruction Set Computer}
\newacronym{riscv}{RISC-V}{Reduced Instruction Set Computing - V}
\newacronym{rtl}{RTL}{Register-Transfer Level}
\newacronym{sdk}{SDK}{Software Development Kit}
\newacronym{seec}{SEEC}{SElf-awarE Computing}
\newacronym{sgpr}{SGPR}{Scalar General Purpose Register}
\newacronym{si}{SI}{Southern Island}
\newacronym{simd}{SIMD}{Single Instruction Multiple Data}
\newacronym{simf}{SIMF}{Single Instruction Multiple Floating point}
\newacronym{sm}{SM}{Streaming Multiprocessor}
\newacronym{snr}{SNR}{Signal to Noise Ratio}
\newacronym[plural=SoCs,firstplural=Systems on Chip (SoCs)]{soc}{SoC}{System-on-Chip}
\newacronym{spared}{SPARED}{Self-aware PArtial Reconfiguration architecture for Edge Devices}
\newacronym{spice}{SPICE}{Simulation Program With Integrated Circuit Emphasis}
\newacronym{tad}{TAD}{Time \gls{adc}}
\newacronym[plural=TDCs,longplural={Time-to-Digital Converters}]{tdc}{TDC}{Time-to-Digital Converter}
\newacronym{tq}{TQ}{Time-Quantizer}
\newacronym{uart}{UART}{Universal Asynchronous Receiver/Transmitter}
\newacronym{vcdu}{VCDU}{Voltage Controlled Delay Unit}
\newacronym{vco}{VCO}{Voltage Controlled Oscillator}
\newacronym{vga}{VGA}{Video Graphics Array}
\newacronym{vhdl}{VHDL}{Very High Speed Integrated Circuit Hardware Description Language}
\newacronym{vlsi}{VLSI}{Very Large Scale Integration}
\newacronym{vgpr}{VGPR}{Vector General Purpose Register}
\newacronym{xilffs}{XILFFS}{Generic Fat File System Library}

\newacronym{amd}{AMD}{Advanced Micro Devices}
\newacronym{beol}{BEOL}{Back End Of Line}
\newacronym{cad}{CAD}{Computer-Aided Design}
\newacronym{cas}{CAS}{Circuits and Systems}
\newacronym{dfa}{DFA}{Discriminant Function Analysis} 
\newacronym{eu}{EU}{European Union}
\newacronym{fdd}{FDD}{fault detection and diagnostic}
\newacronym{fefet}{FeFET}{Ferroelectric Field Effect Transistor}
\newacronym{feline}{FeLINe}{FeFET Logic IN mEmory}
\newacronym[plural=FoMs,longplural={Figures of Merit}]{fom}{FoM}{Figure of Merit}
\newacronym{ga}{GA}{Genetic Algorithm}
\newacronym{hipeac}{HiPEAC}{High Performance and Embedded Architecture and Compilation}
\newacronym{hp}{HP}{Hewlett Packard}
\newacronym{hqp}{HQP}{Highly Qualified People}
\newacronym{hvac}{HVAC}{Heating, Ventilation and Air Conditioning}
\newacronym{ibm}{IBM}{International Business Machines corporation}
\newacronym{ict}{ICT}{Institute for Computer Technology}
\newacronym{iot}{IoT}{Internet of Things}
\newacronym{nda}{NDA}{Non-Disclosure Agreement}
\newacronym{nvp}{NVP}{Non-Volatile Processor}
\newacronym{oecd}{OECD}{Organization for Economic Cooperation and Development}
\newacronym{rd}{R\&D}{Research and Development}
\newacronym{rmsdd}{RMSDD}{Root-Mean Square of Successive Differences}
\newacronym{sdnn}{SDNN}{Standard Deviation Normal-to-Normal-Intervals}
\newacronym{soa}{SoA}{State-of-the-Art}
\newacronym{tsmc}{TSMC}{Taiwan Semiconductor Manufacturing Company}
\newacronym{tvlsi}{TVLSI}{Transactions on Very Large Scale Integration}

\newacronym{felix}{FELIX}{Fast and energy-efficient Logic in Memory}
\newacronym{esd}{ESD}{Electrostatic Discharge}
\newacronym{sdc}{SDC}{Self Directed Channel}
\newacronym{tmsl}{TMSL}{Three Memristors Stateful Logic}

\begin{document}
\title{BEhavioral Leakage and IntEr-cycle 
Variability Emulator model for ReRAMs (BELIEVER)}
\author{David Radakovits and Nima Taherinejad 
\thanks{D. Radakovits and N. TaheriNejad are with the TU~Wien, 1040 Vienna, Austria}
\thanks{This work has been submitted to the IEEE for possible publication. Copyright may be transferred without notice, after which this version may no longer be accessible.}
}

\maketitle

\begin{abstract}
Emerging electronic devices are promising to drive the performance of computer systems to new heights, against the notable saturation in traditional transistor-based architectures. Among them, resistive RAM -- or ReRAM -- has attracted a lot of attention among scientists since its practical realization was reported in 2008 and numerous devices, circuits and systems, and also models have been described in the literature. However, behavioral models fail to reproduce device parameter variations and the drift of device state in the absence of a stimulus. This shortcoming substantially reduces the practical relevance of systems and circuits designed with existing models. The work at hand deals with the development of a behavioral model that integrates device parameter variation and state drift
based on data collected from our measurements of real devices. As we show in this paper, BELIEVER model enables engineers to conduct more reliable and meaningful design and simulations of circuits and systems that use ReRAMs. 
\end{abstract}

\glsresetall

\section{Introduction}
\label{chap:introduction}

Requirements for computer systems are in an ever increasing growth, especially today, where Big Data applications \cite{madden2012databases,sagiroglu2013big,chen2014data,chen2014big} on the one hand and the \gls{iot} on the other hand span a wide spectrum of requirements from high performance to low energy consumption and footprint. Interestingly both ends of the spectrum are still served by computer systems that are more or less based on the same architecture, namely the Von Neumann-architecture. 
{A well-established technique to alleviate the Von Neumann-bottleneck is caching, where memory elements are located near the \gls{cpu}. \gls{imc} goes the other way round, i.e., installing  elements in the memory, which are capable of doing logic operations, without the data leaving the memory~\cite{linn2012beyond,li2016pinatubo,Hur2016,gaillardon2016programmable,Papandroulidakis2017,li2018memory}.}

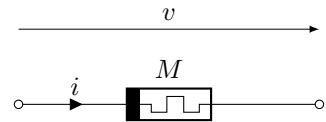
\begin{figure}[]
    \centering


\def\xlen{4mm}

    \begin{circuitikz}
        \draw (5,1) node(right1){} to[memristor, l_=$M$, o-o]+(-4,0) node(left1){};
        \draw (1.7,1) to[short,i=$i$] (1.8,1);
        
        \node at (3,2.2) {$v$};
       \draw[-latex,line width=0.3pt] (1,2) -- +(4,0);
    \end{circuitikz}
    \caption[Memristor symbol with reference directions]{Memristor symbol with reference direction for the current  and voltage. }
    \label{fig:memristorsymbol}
    \vspace{-3mm}
\end{figure}

The memristor is a basic element in electrical circuits, along with resistor, capacitor and inductor. Its theoretical behavior was first described in 1971 by Leon Chua \cite{chua1971memristor}
{and in}
2008 the first passive realization of a device that behaves like a memristor was reported~\cite{strukov2008missing}. 
Although memristive effects have been observed before without recognizing them as such \cite{Menzel2018,Rainer2009}, this device was the first to intentionally behave like a memristor in terms of voltage-current relationship over time.
The memristors that have been studied in this work,  are of the \gls{sdc} type~\cite{knowm2020sdc}, which are called resistance switching devices or {ReRAMs}~\cite{gale2014tio2}. The {aforementioned} memristor reported in~\cite{strukov2008missing} and many others~\cite{lohn2013optimizing,jin2014study,jiang2016sub,abunahla2018switching,lian2017characteristics} are also of the {ReRAM} type. Although in the community the term \textit{memristor} is established as an umbrella term for different kinds of memristive devices apart form {ReRAM}s, e.g., \gls{stt} or \gls{pcm}~\cite{meena2014overview,rizk2019demystifying}, whenever the term is used in this work, it refers to \textit{resistance switching device}.
The commonly used memristor symbol and its reference directions in terms of current and resistance change is shown in \Cref{fig:memristorsymbol}.

\begin{figure}[b]
\vspace{-3mm}
    \centering
    \begin{tikzpicture}
  \foreach \percent/\name/\color in {
    88/Other Works/light blue,
  } {
    \ifx\percent\empty\else                 
    \global\advance\ind by 1              
    \draw[fill={\color!100},draw={\color}] (0,0) -- (\angle:\radius) arc (\angle:\angle+\percent*3.6:\radius) -- cycle;
    \node[anchor=base west] at (\radius+0.25,0.5-.5*\ind) {\squaretwo{\color} \name}; 
    \node[rectangle]() at (\angle+0.5*\percent*3.6:0.5*\radius) {\percent\%}; 
    \pgfmathparse{\angle+\percent*3.6}    
    \xdef\angle{\pgfmathresult}           
    \fi
  };
  \foreach \percent/\name/\color in {
    12/Circuit{\slash}System Implementation/light red,
  } {
    \ifx\percent\empty\else                 
    \global\advance\ind by 1              
    \draw[fill={\color!100},draw={\color}] (0,0) -- (\angle:\radius) arc (\angle:\angle+\percent*3.6:\radius) -- cycle;
    \node[anchor=base west] at (\radius+0.25,0.5-.5*\ind) {\squaretwo{\color} \name}; 
    \node[rectangle]() at (\angle-0.1*\percent*3.6:2*\radius) {\percent\%}; 
    \pgfmathparse{\angle+\percent*3.6}    
    \xdef\angle{\pgfmathresult}           
    \fi
  };
  \draw (.8,2.1) -- (1.05,2.7) -- +(1.15,0);
\end{tikzpicture}
    \caption[Share of practically implemented and measured memristive circuits and systems]{Share of practically implemented and measured memristive circuits and systems from 
    Google Scholar search results on memristors, {ReRAM}, resistance switch and other related keywords
    ~\cite{Taherinejad2019cas}.}
    \label{fig:pie}
\end{figure}
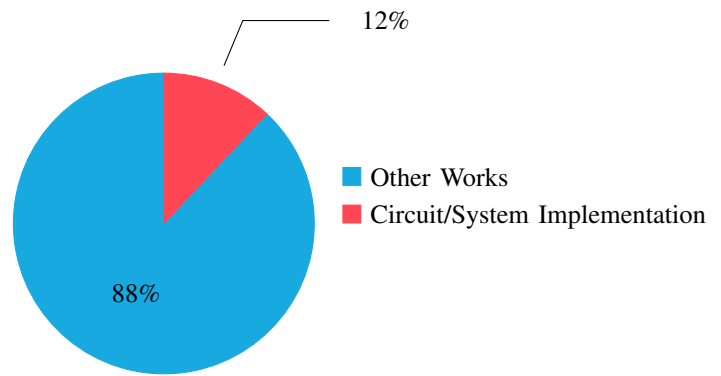

An important step in hardware development is simulation. Models and modeling techniques such as \gls{spice} have evolved in the last decades and make simulation and design of complex hardware possible throughout different levels of abstraction. While modeling of well-explored technologies such as \gls{cmos} has gained a level, in which engineers can rely on simulation results with very little probability of false predictions, this is not the case for emerging technologies such as memristors. As {it} was shown in~\cite{Taherinejad2019cas}, simulations of memristive circuits and systems are not backed up by practical implementations and measurements in a vast portion of cases. \Cref{fig:pie}, which was adopted from~\cite{Taherinejad2019cas}, illustrates, that only 12\% of the top 30 Google Scholar search results on `memristor', `{ReRAM}', `resistance switch' and other keywords, actually report implementations and measurements of memristive circuits and systems.

This circumstance is quite alarming, since the available models do not reproduce cycle-to-cycle or device-to-device variations. 
{Leakage,}
i.e., a change of state in absence of a stimulus,
{is not} 
reproduced in 
models. While these mostly unwanted effects are well-explored in traditional technologies and their respective models (such as \gls{cmos}), the memristor community still needs to close this gap. 
To increase the simulation reliability of memristive circuits and systems, it is important to reproduce such effects in models.

Several memristor models are available in the literature.
Generally they can be divided into two major groups: Those which model memristor behavior on a physical or electro-chemical level, e.g., \cite{jiang2014verilog}, and those which model macroscopic voltage-current behavior on a high level of abstraction, e.g., \cite{kvatinsky2015vteam,yakopcic2011memristor,biolek2013reliable}. While the latter group naturally lacks exact reproduction of memristor internals, it has advantages, such as lower computational cost and a certain flexibility to be used among devices with similar functional internal behaviors (e.g., different resistance switching devices), given appropriate model fitting. Since there is no physical or chemical model available for the memristors investigated in this work and considering the mentioned advantages, the latter group was examined. However, to the best of the authors knowledge, none of the models reproduce device variability or leakage effects. 
Therefore, this work shall accomplish two major tasks: Fitting a memristor model, that is well received by the memristor community to \gls{sdc} memristor devices built by KNOWM Inc., as well as incorporating variation and leakage effects into the mentioned model. 

The designed experiments, the forming of the memristor under test and the outcome of the conducted experiments are clarified in \Cref{chap:experiments}. The base model that was chosen to build the enhanced model upon, the model development process and the fitting of the enhanced model are explained in detail in \Cref{chap:model}. 
The impact of the observed and modeled effects on popular memristive logics is showcased in \Cref{chap:casestudy}.
The work at hand is concluded and an outlook to potential future works is given in \Cref{chap:conclusion}.

\section{Experiments}
\label{chap:experiments}

The extraction of the necessary memristor parameters demands several different experiments to be conducted. The purpose, design and outcome of each experiment is explained in this chapter.

\subsection{Forming}
\label{sec:forming}
\begin{figure}
    \centering
    \includegraphics[width=0.9\linewidth]{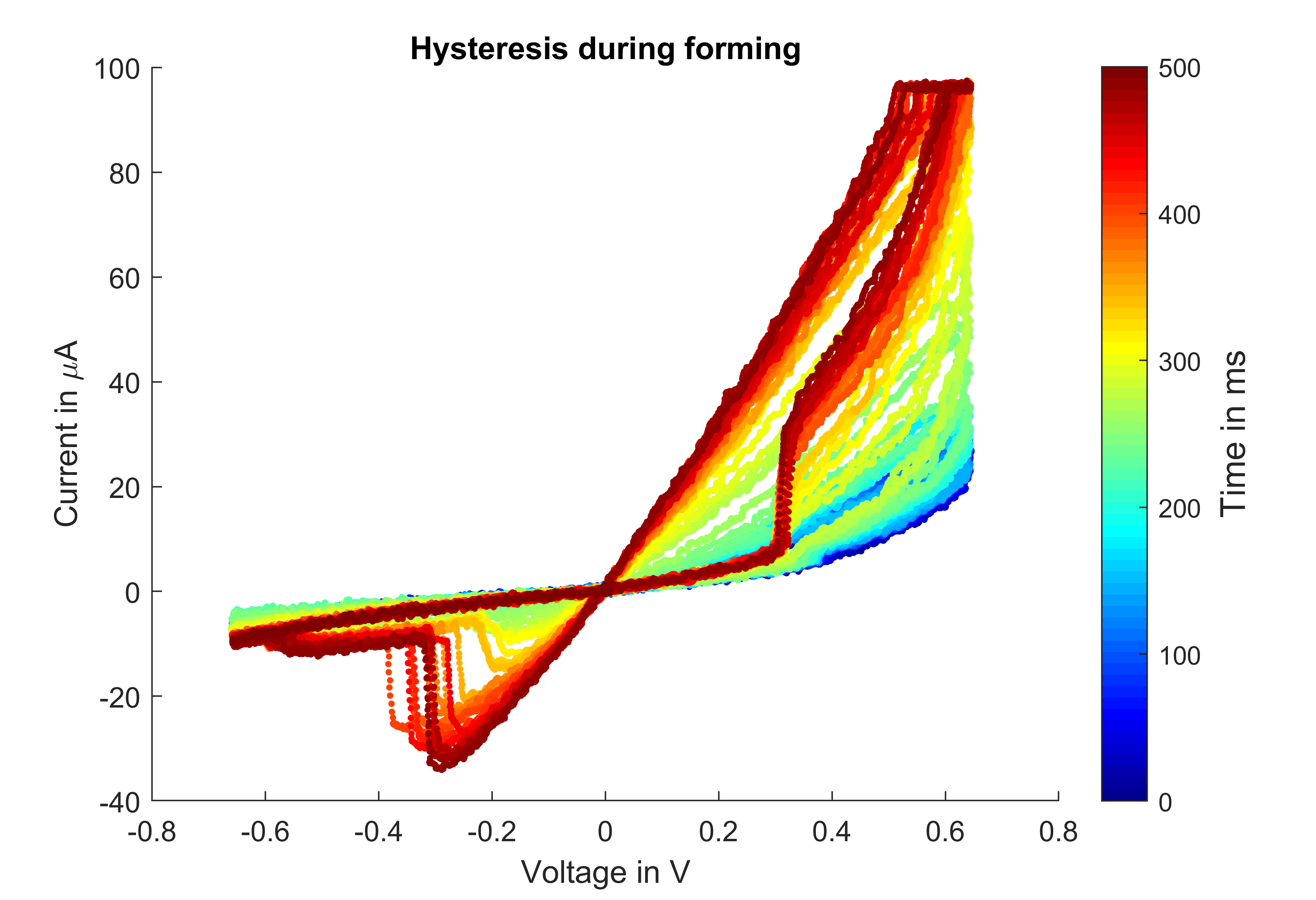}
    \caption[Hysteresis during forming]{Evolution of the voltage-current curve (hysteresis) of the memristor during forming.  
    The color of the data points indicates the time, i.e., blue indicates early measurement points, red indicates late points. }
    \label{fig:forming_evolution}
\end{figure}

Before an \gls{sdc} memristor can be used,
it has to be \textit{formed}. Forming is the process of initial creation 
of the conducting paths in the filament of the memristor. 
In order to form the device, as recommended by the manufacturer~\cite{knowm2020sdc}, a sinusoidal voltage with a frequency of $100$\,Hz is applied to the memristor. The amplitude is gradually increased until the typical hysteresis can be observed. 
\Cref{fig:forming_evolution} shows the hysteresis of the memristor, which is the evolution of the voltage-current curve of the memristor, during forming. 
The color of the data points indicates the time, i.e., blue indicates early measurement points, red indicates late points. 
As {it} can be seen from the flattening of the curve in the upper right corner of the figure, the forming, as all other experiments, was conducted with a current limit set to approximately $100$\,$\mu$A. The data sheet for the \gls{sdc} memristors provided by the manufacturer~\cite{knowm2020sdc} gives a maximum current of $1$\,mA. However, because Gomez~et~al. used $800$\,$\mu$A with the same device in~\cite{gomez2019exploring} and reported a very short life time for the devices, this large safety margin was chosen.
Note that the current limit during forming has a large impact on memristor parameters, especially, but 
not exclusively on the \gls{hrs} and \gls{lrs}
resistance. As {it} can be seen in the memristor datasheet~\cite{knowm2020sdc}, \gls{hrs} resistance, \gls{lrs} resistance and also the span of the two decreases with increasing current limit during forming process.

\subsection{Resistance Variation}
\label{sec:exp_RON_ROFF}
\begin{figure}[!t]
    \centering
    \includegraphics[width=\linewidth]{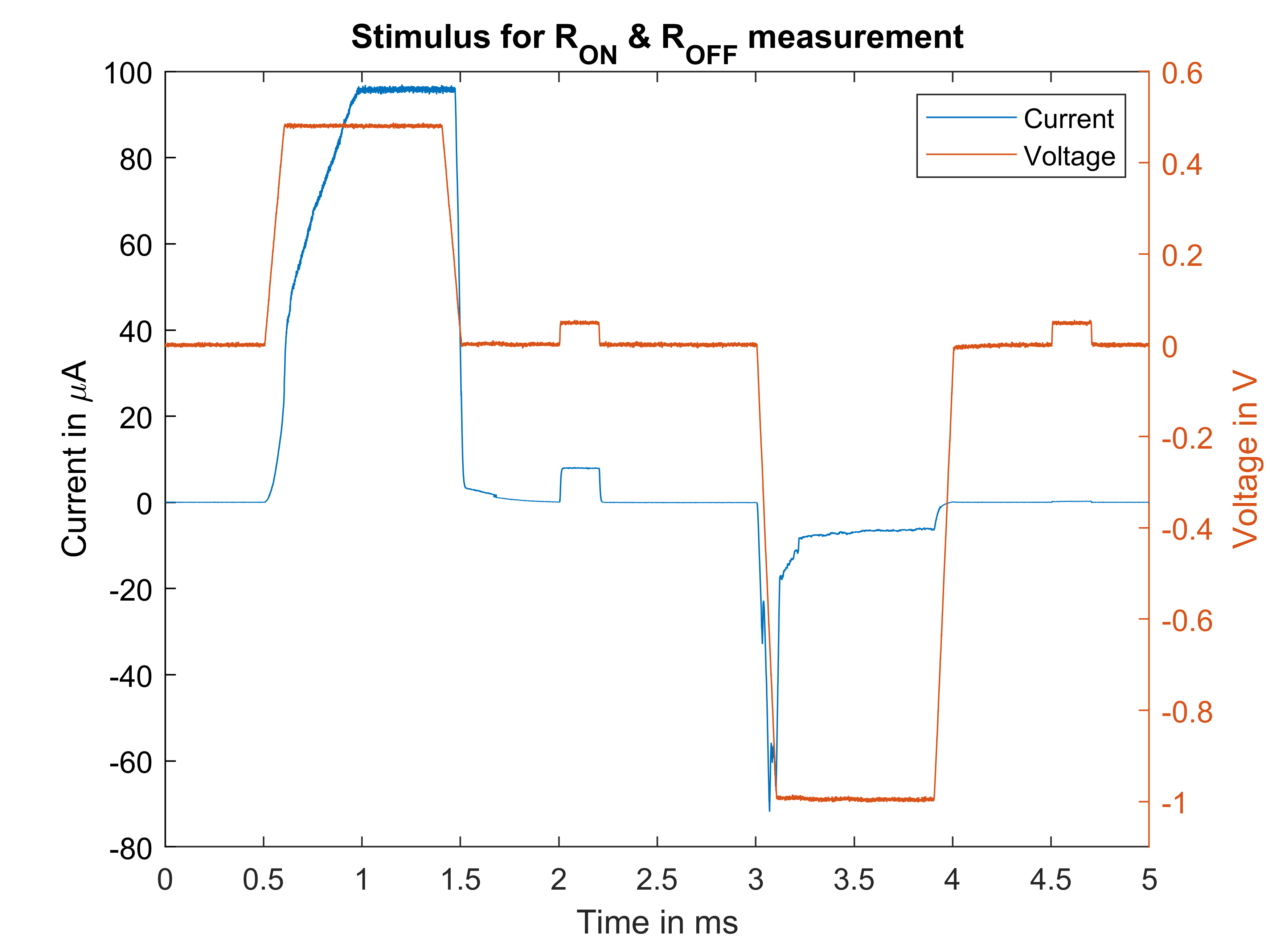}
    \caption[SET and RESET stimulus for measurement of \Ron{} and \Roff{} variation]{SET and RESET stimulus for measurement of \Ron{} and \Roff{} variation. The orange graph denotes the voltage that was applied to the memristor and the blue graph denotes the memristor current. 
    }
    \label{fig:RON_ROFF_curve}
\end{figure}

The variation of the resistance that is measured after the SET(RESET) stimulus, i.e., the stimulus that is supposed to drive the memristor into \gls{lrs}(\gls{hrs}), is of great interest for the designers of a memristive circuit, since in most cases the circuit will need to be tuned to that parameter. In order to measure the cycle-to-cycle variation of the resistance after SET (\Ron{}) and RESET (\Roff{}), 
the stimulus in~\Cref{fig:RON_ROFF_curve} was applied 100 times to the memristor under test. As {it} can be seen from the orange graph in the figure, denoting the applied voltage, first a $1$\,ms wide $500$\,mV SET pulse was used to drive the memristor into \gls{lrs}. The drop of the memristor's resistance can be seen in the increasing current in the blue graph. The resistance of the memristor after the SET pulse was determined during the $200$\,$\mu$s wide $50$\,mV measurement pulse that was applied afterwards. After measuring \Ron{} this way, a 1ms wide\, $-1$\,V RESET pulse was applied to the memristor in order to drive it to \gls{hrs} and the resistance in \gls{hrs} was measured with a $200$\,$\mu$s wide $50$\,mV measurement pulse again. Theoretically the resistance of the memristor in \gls{lrs}(\gls{hrs}) could have been measured at the end of each SET(RESET) pulse. This would however hide a potential non-linear voltage-current relationship. If the memristor behaved like an ideal, i.e. linear, resistor while the state is constant, any given measurement voltage would result in the same resistance. Since it is not known, if the memristor does behave linearly, \Ron{} and \Roff{} have to be measured using the same voltage. The measurement voltage of 50mV was chosen upon the experience gained during forming: As {it} can be seen in~\Cref{fig:forming_evolution} the resistance starts to change significantly around $300$\,mV during SET and 
{around}\, 
$-150$\,mV during RESET. Any voltage in between that range would theoretically qualify as a measurement voltage. Too small measurement voltages, however, lead to very low currents, which decrease the \gls{snr} of the measurement. Note, that the plateau in the current during SET in \Cref{fig:RON_ROFF_curve} does not indicate that the resistance of the memristor does not change anymore. The current saturation is caused by the current limit set to approximately $100$\,$\mu$A. However, in \Cref{sec:exp_res_change} it is shown that, at the used amplitude, the resistance change saturates after approximately $1$\,ms, making the described stimulus suitable to drive the memristor into full \gls{lrs}. During RESET the current saturates in the first half of the applied pulse, ensuring that the memristor is fully driven into \gls{hrs}.

\begin{figure}[t!]
    \centering
    \includegraphics[width=\linewidth]{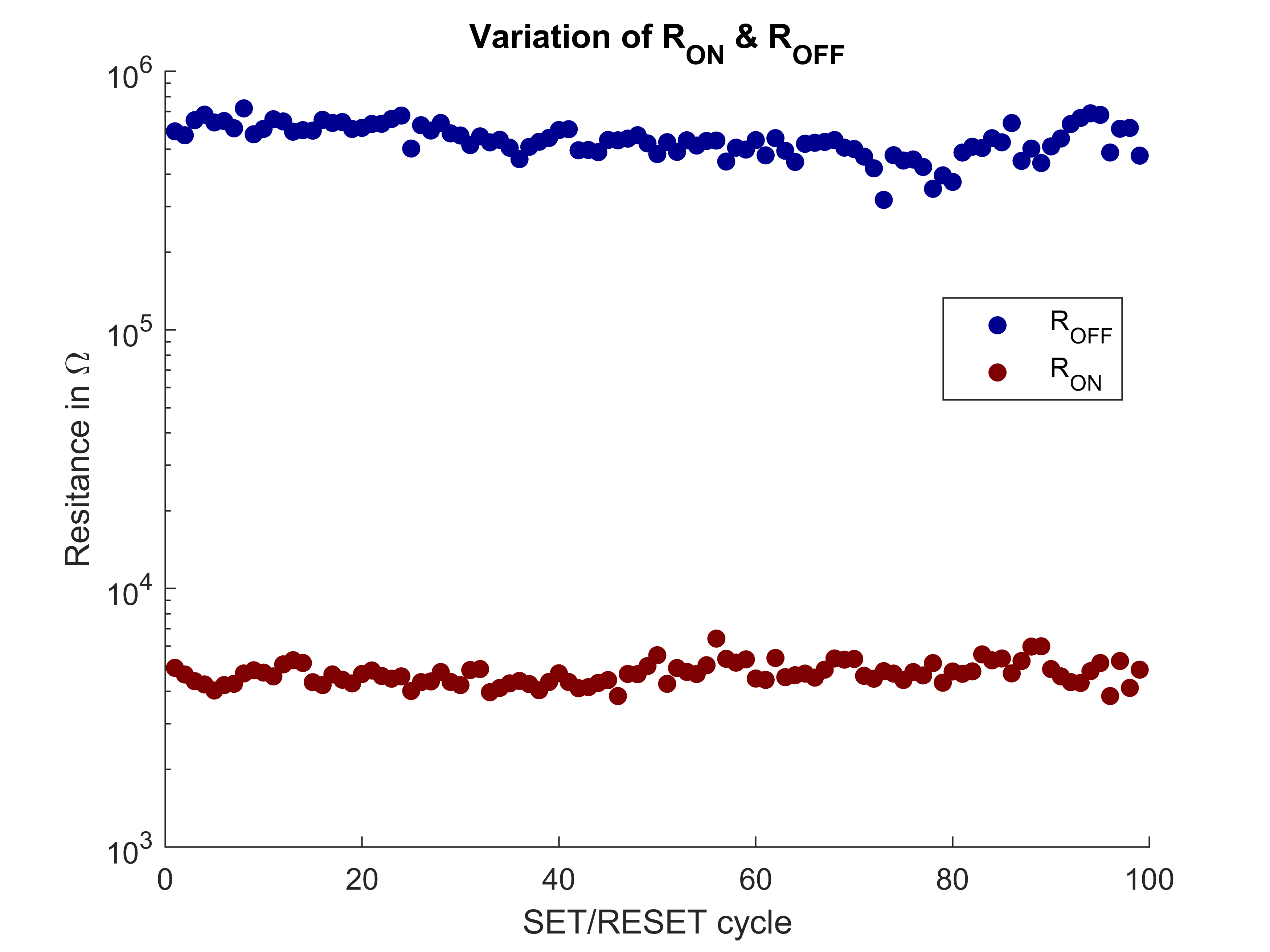}
    \caption[Example of variation of \Ron{} after SET and \Roff{} after RESET stimulus]{Example of variation of \Ron{} after SET and \Roff{} after RESET stimulus. Blue dots denote the measured resistance after a RESET stimulus, red dots denote the measured resistance after the SET following the RESET stimulus.}
    \label{fig:RON_ROFF_scatter}
\end{figure}

\begin{figure}
    \centering
    \includegraphics[width=0.8\linewidth]{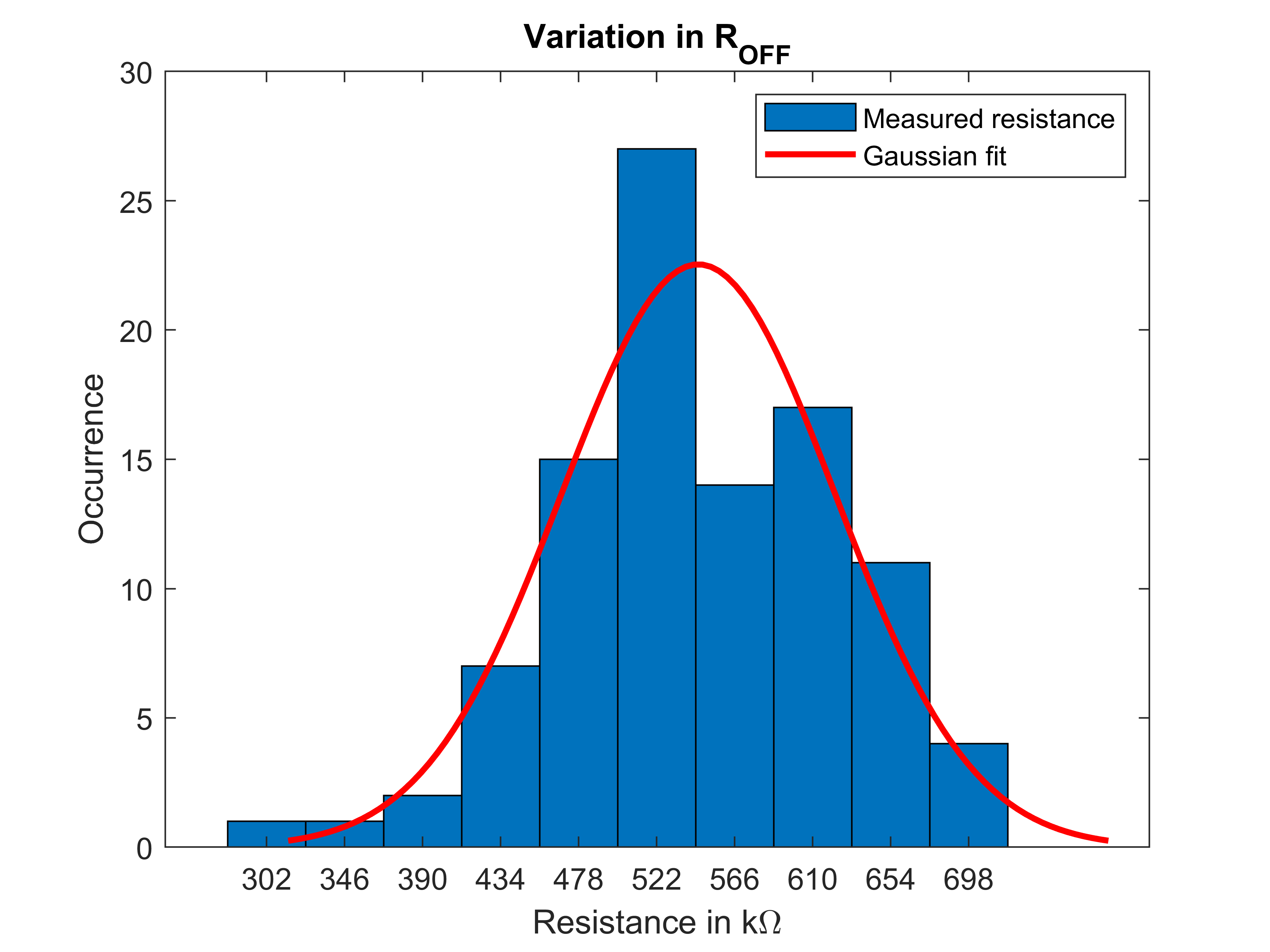}
    \caption[Histogram of measured \Roff{} values and fitted normal distribution]{Histogram of measured \Roff{} values and fitted normal distribution: $\mu=545.52$\,k$\Omega$, $\sigma=77.065$\,k$\Omega$.}
    \label{fig:ROFF_hist}
\end{figure}

\begin{figure}
    \centering
    \includegraphics[width=0.8\linewidth]{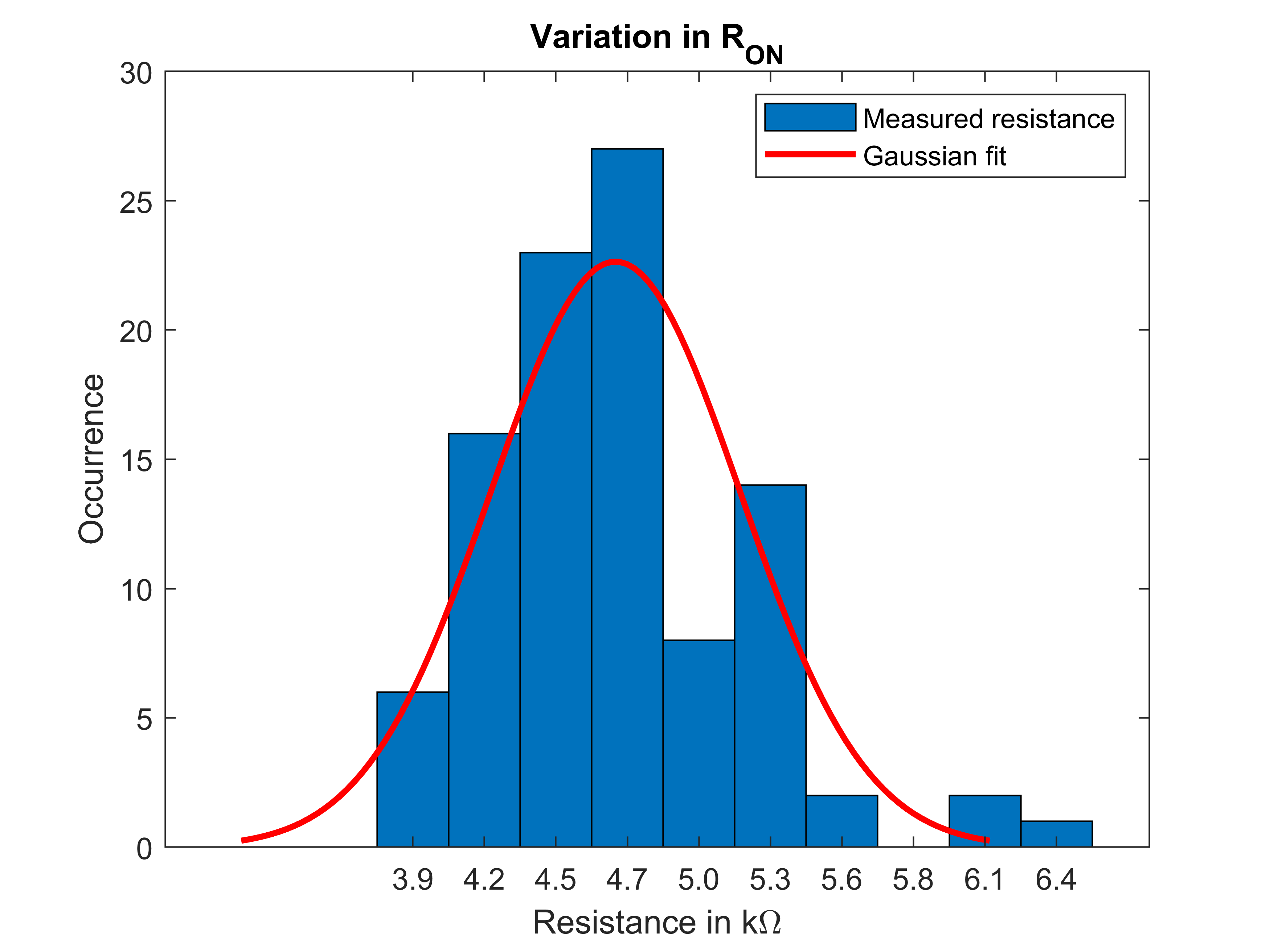}
    \caption[Histogram of measured \Ron{} values and fitted normal distribution]{Histogram of measured \Ron{} values and fitted normal distribution: $\mu=4.64$\,k$\Omega$, $\sigma=427.9$\,$\Omega$.}
    \label{fig:RON_hist}
\end{figure}

\Cref{fig:RON_ROFF_scatter} 
shows an example of the conducted experiment. As {it} can be seen
in the figure, the resistance in both \Ron{} and \Roff{} varies around a mean value. The mean value of \Ron{} (depicted in the figure in red) is $4.64$\,k$\Omega$, the mean value of \Roff{} (depicted in the figure in blue) is $545.52$\,k$\Omega$.
Figures~\ref{fig:ROFF_hist}~and~\ref{fig:RON_hist}
show the histogram of the recorded values of \Roff{} and \Ron{}, respectively. The standard deviation is  $\sigma=427.9$\,$\Omega$ for \Ron{} and $\sigma=77.095$\,k$\Omega$ for \Roff{}. 
A normal or gaussian distribution was chosen for both \gls{hrs} and \gls{lrs} for homogeneous and thus more convenient implementation of the variation model.

\subsection{Resistance Change Dynamics}
\label{sec:exp_res_change}
\begin{figure}[t!]
     \centering
     \includegraphics[width=\linewidth]{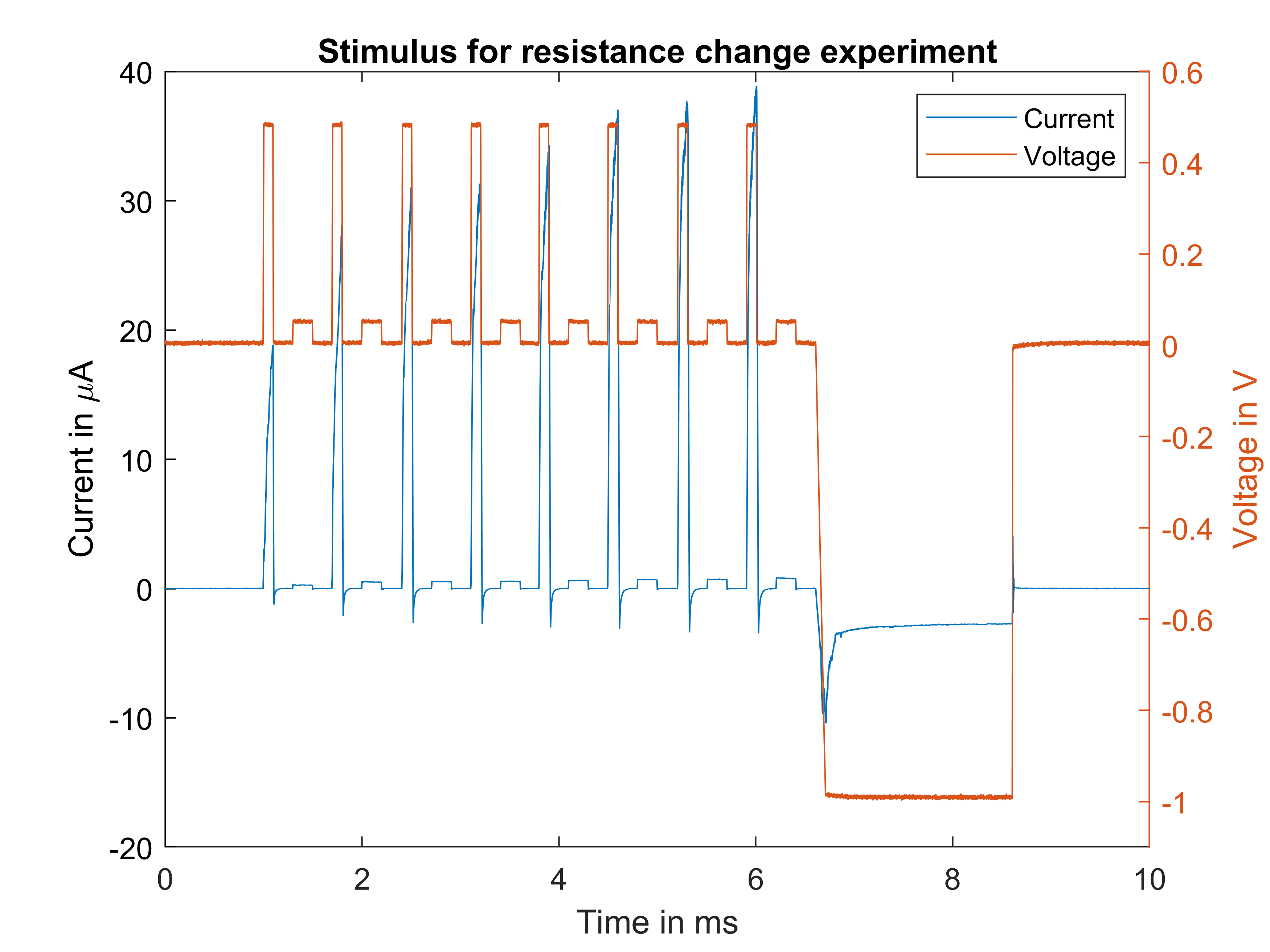}
     \caption[Stimulus applied for the measurement of resistance change]{Stimulus applied for the measurement of resistance change. Eight pairs of 500\,mV$\times$\,$100$\,$\mu$s SET pulse and 50\,mV$\times$\,$200$\,$\mu$s measurement pulse are followed by a\, $-1$\,V$\times$\,$2$\,ms RESET pulse.}
     \label{fig:res_change_curve}
\end{figure}

\begin{figure}
\centering
\subfigure[Resistance after each applied SET pulse]{
     \label{fig:res_change_all}
    \includegraphics[width=0.46\linewidth]{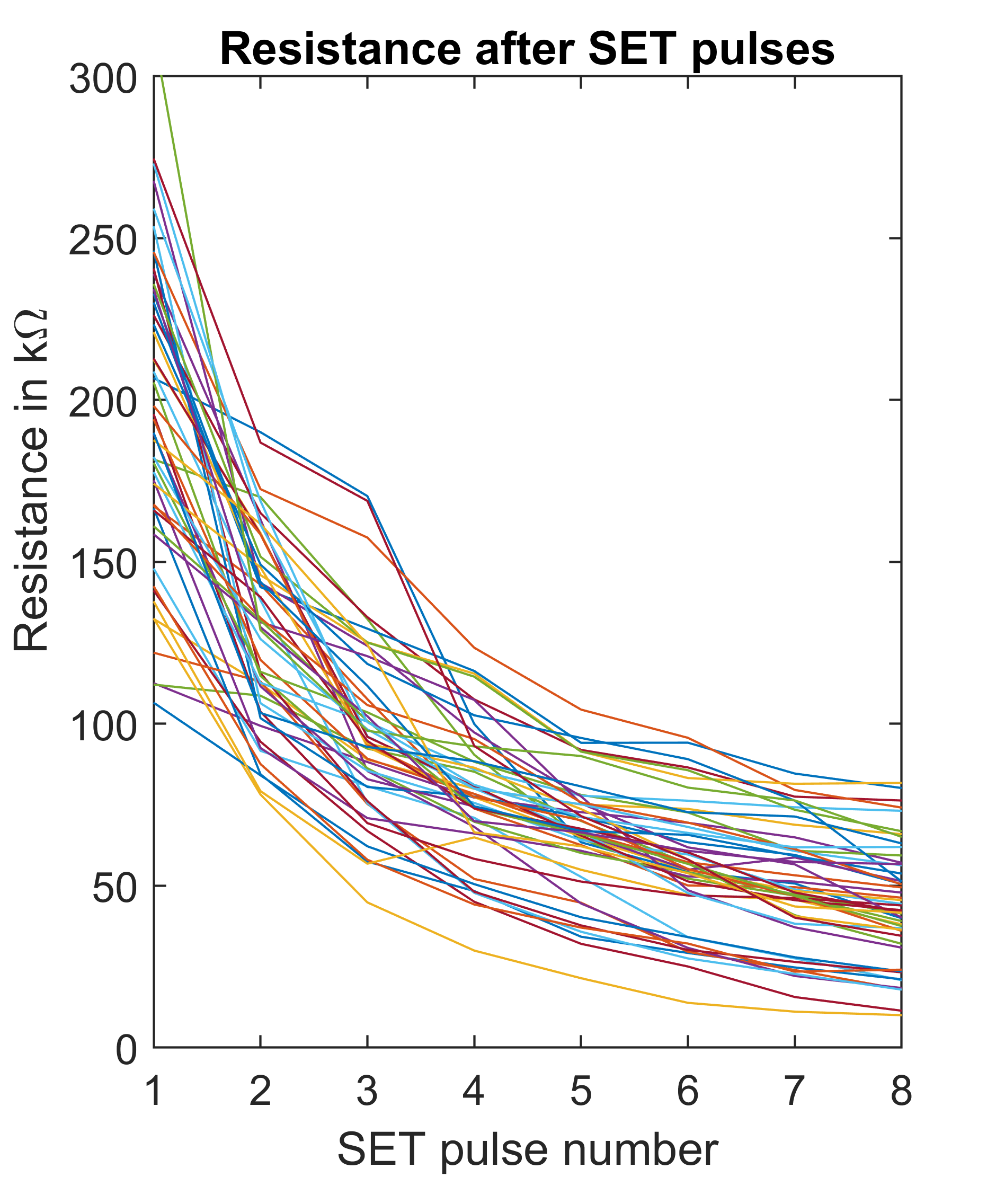}
    }
\subfigure[Evolution of resistance during RESET stimulus]{
    \label{fig:RESET_time}
    \includegraphics[width=0.44\linewidth]{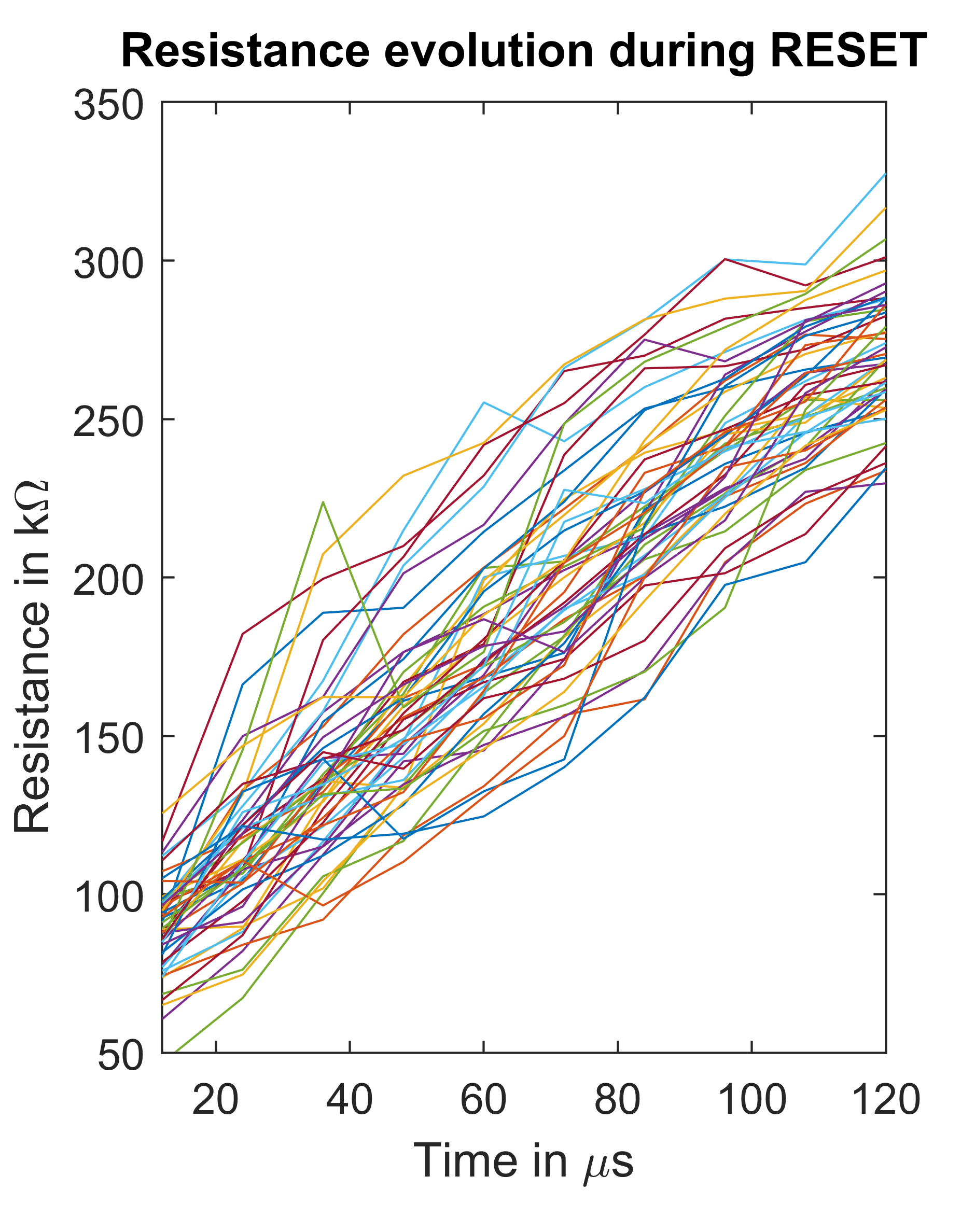}
    }
\caption[Evolution of resistance]{Evolution of resistance during SET and RESET.
}
\end{figure}

To determine the speed of state change during the SET process, an experiment was conducted repeatedly, that consisted of a series of eight SET pulses and a RESET pulse. Each SET and RESET pulse was followed by a measurement pulse during which the  
resistance of the memristor at that instance was determined. The SET pulses measured $500$\,mV$\times$\,$100$\,$\mu$s, the measurement pulses measured $50$\,mV$\times$\,$200$\,$\mu$s and the RESET pulse measured $-1$\,V$\times$\,$2$\,ms
{, as it can be seen in} \Cref{fig:res_change_curve}.
\Cref{fig:res_change_all} shows the recorded resistance after each of the eight applied SET pulses for all $100$ measurements{, in which each line} 
denotes the evolution of the resistance as it is influenced by the eight applied SET pulses.
\Cref{fig:res_change_model} shows the resistance (in blue) after each SET pulse averaged over all 100 measurements. Using this data, the resistance change per applied stimulus of a given model can be fitted to the measured memristor behavior.

The RESET process dynamics could not be examined in the same fashion as the SET process, because it
{was not possible} to find an appropriate stimulus for a similar, pulsed measurement.{ That is, no configuration of pulse length and height with the described setup could be found, that would switch the memristor only partially instead of full \gls{hrs}.} 
Hence, the aforementioned RESET pulse that was used to reset the memristor after the applied series of eight SET pulses was used to 
study the RESET dynamics. \Cref{fig:RESET_time} shows the evolution of the resistance during each individual RESET pulse. \Cref{fig:RESET_change_model} shows the average evolution of resistance (in blue) during all applied RESET pulses.
{It should be noted that,
since the resistance is determined at an applied voltage of\, $-1$V, the RESET process is ongoing at that time, which may distort the measurement.}

Moreover,
the resistance is measured at a different voltage than for the SET process experiment, which may
affect the comparability between the SET and RESET experiment due to non-linear voltage-resistance behavior.

\subsection{Threshold Voltage}
\label{sec:exp_threshold}
\begin{figure}[t!]
    \centering
    \includegraphics[width=0.9\linewidth]{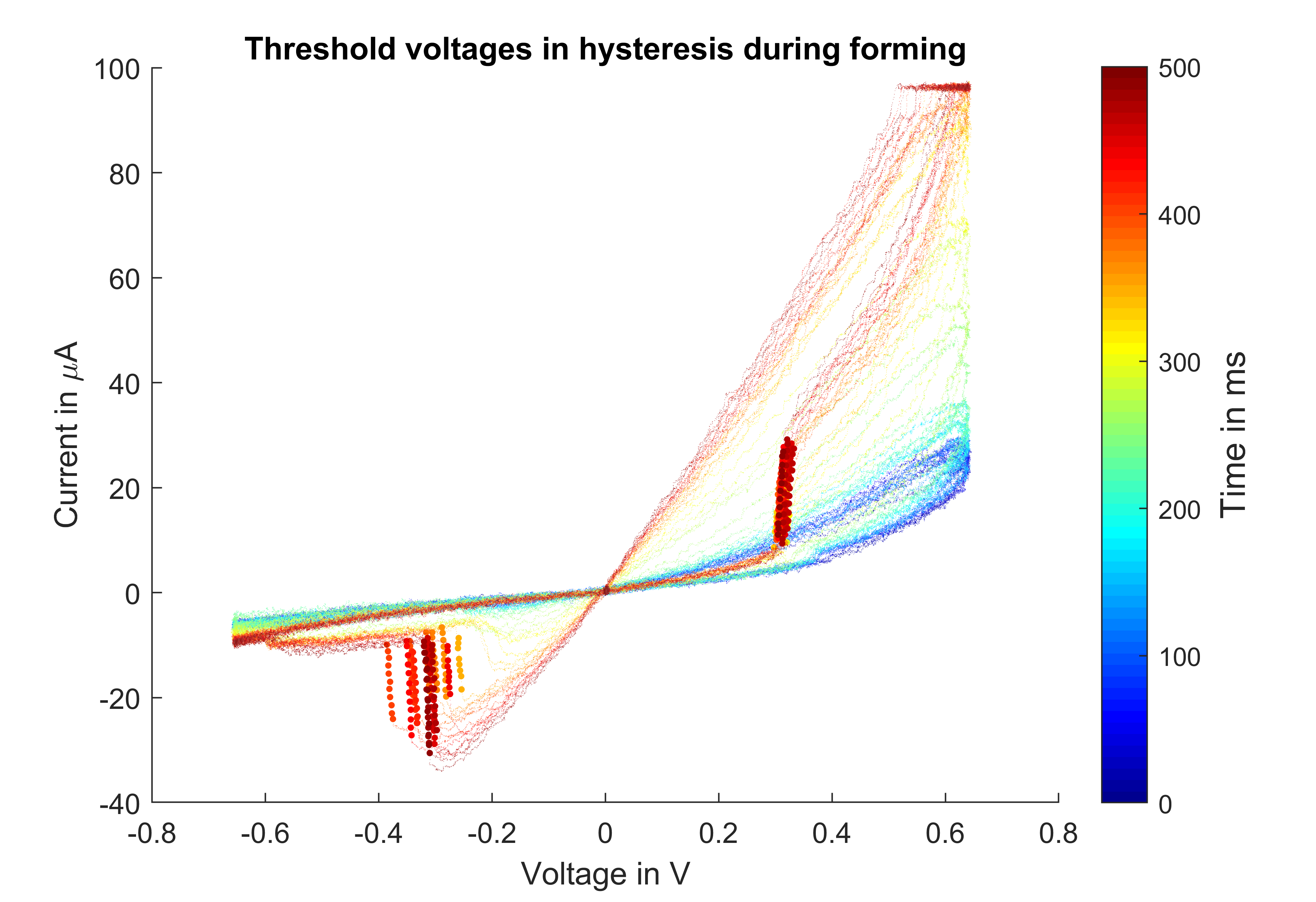}
    \caption[Hysteresis recorded during forming with highlighted areas of threshold effect]{Example of hysteresis recorded during forming (see \Cref{fig:forming_evolution} and \Cref{sec:forming}) with highlighted areas (bold dots near $-300$mV and $+300$mV), used for determining threshold voltages. The color of the dots determines time, i.e., dark blue denote early points, dark red denote late points. 
    }
    \label{fig:hysteresis_thresholds}
\end{figure}

The threshold voltages of the memristor were
extracted from the hystereses recorded during forming of the memristor. In order to do so, the threshold voltages were defined as the voltages, at which a major change in resistance, i.e., a change that is above a defined threshold, happens. This threshold was empirically found to be $\Delta I=2$\,A/s 
$\approx$ 80\,nA/sample. A \textit{{MATLAB}} back-end for the {\textit{LabVIEW}} environment was developed to automatically derive 
the threshold voltages from the recorded hystereses. 
\Cref{fig:hysteresis_thresholds} shows an example of a recorded hysteresis (the same hysteresis as in \Cref{fig:forming_evolution} was used as the example) with the highlighted areas of threshold effects, which have been identified by the software. 
The average threshold voltages of all recorded hystereses are $V_\mathrm{TH+}=370.2$\,mV and $V_\mathrm{TH-}$=$-373.8$\,mV. Fifty test cycles were used to  
{extract}
{threshold voltage values.}

\subsection{Leakage}
\label{sec:exp_leakage}
Leakage refers to a change of the memristor's state represented by the measured resistance change 
at the absence of any stimulus, i.e., without any applied voltage to the memristor or any  
current going through the memristor. This property of a memristor is especially interesting for all kinds of memory and \gls{imc} applications, since it determines when and how the values might be lost if they are not refreshed. 

\begin{figure}
    \centering
    \includegraphics[width=\linewidth]{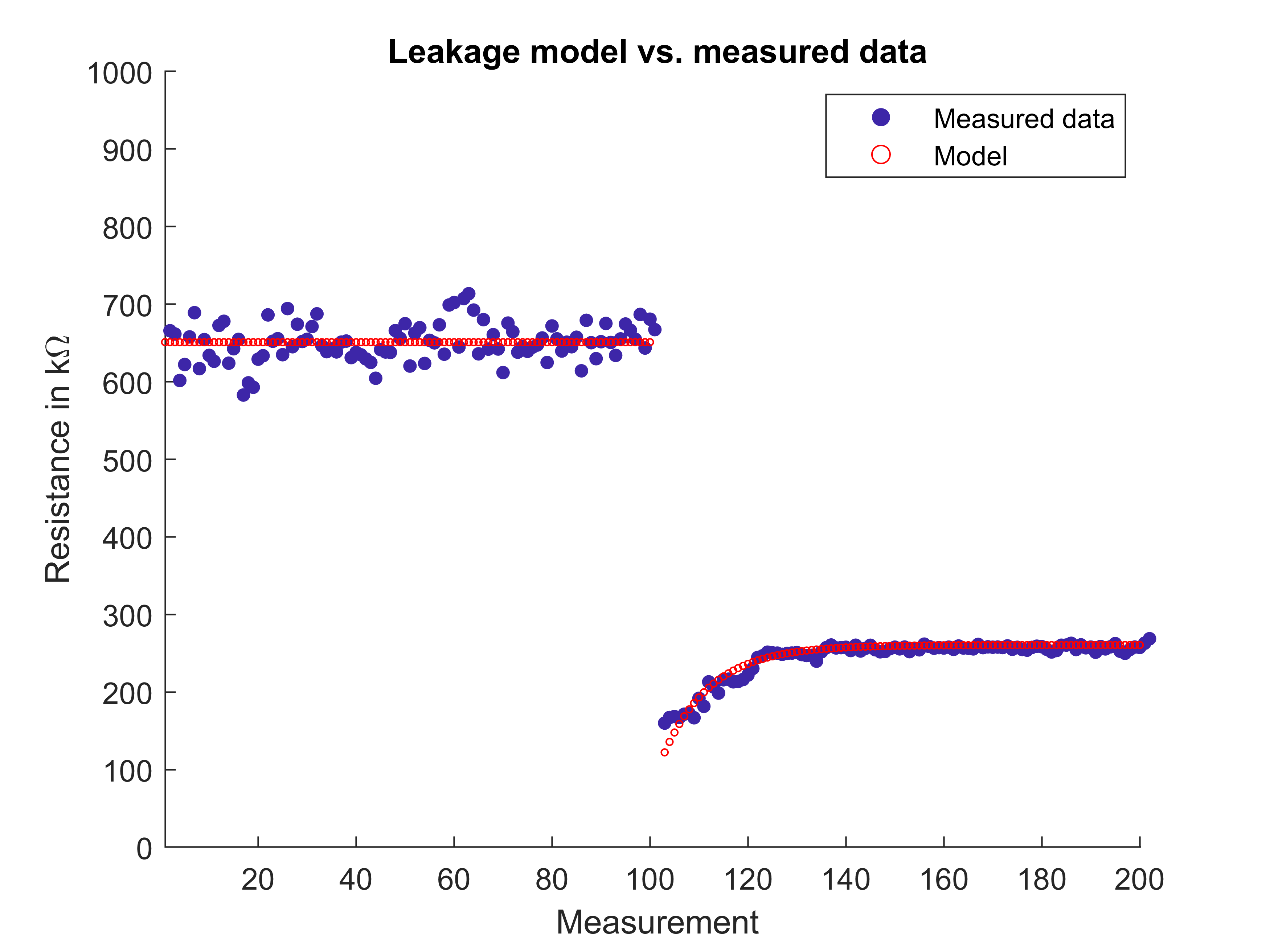}
    \caption[Leakage effect, measurement vs. model]{{Measured leakage effect versus model. Blue: Example of resistance recorded during leakage experiment. $100$ measurements of the resistance were recorded after an applied RESET stimulus (measurements $1$ to $100$) and another $100$ measurements were taken after application of a SET stimulus (measurements $101$ to $200$). The recorded points are $1$\,s apart. Red: Fitted model.}}
    \label{fig:retention_1}
\end{figure}

\begin{figure}
    \centering
    \includegraphics[width=\linewidth]{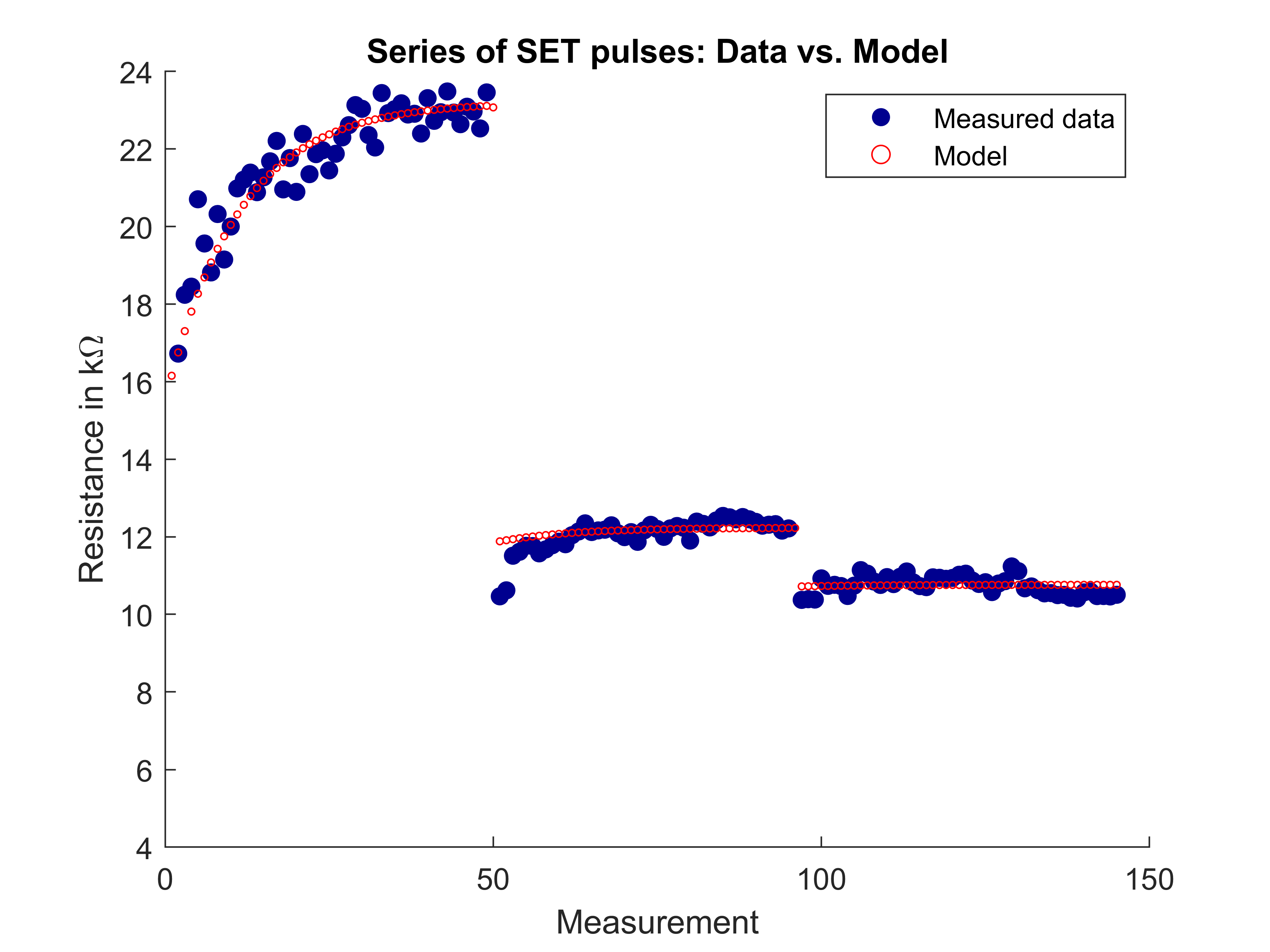}
    \caption[Measured leakage of SET series vs. model]{Resistance during leakage experiment consisting of a series of SET pulses: Measurement versus model. Resistance was measured in $1$\,s time intervals.}
    \label{fig:retention_1_series}
    \vspace{-5mm}
\end{figure}

To measure the leakage of the memristor, the memristor was driven to \gls{hrs}. After that, a series of $100$ measurement pulses ($50$\,mV$\times$\,$200$\,$\mu$s) was applied in 
one second intervals. Then, a SET pulse was applied to the memristor and the resistance was again determined with a series of measurement pulses. \Cref{fig:retention_1} shows the recorded data points. The individual points were recorded in $1$\,s intervals. 
As {it} can be seen from the figure, there is  
no significant state drift  
after RESET. However, there is a state drift present after the SET pulse, which is stronger right after the application of the SET pulse and fades out in approximately $80$-$100$\,s. This experiment was repeated $50$ times. During these tests no considerable state drift after RESET could be  
observed.
The state drift after SET varies in terms of strength, i.e., initial drift rate and duration. However, the general shape that can be seen in \Cref{fig:retention_1} stays present in all experiments. Additionally, an experiment was conducted in which a series of SET pulses was applied after a RESET pulse and the state drift was again recorded in 1s intervals between the SET pulses. \Cref{fig:retention_1_series} shows an example of the experiment, which shows, that state drift is stronger after the first SET pulse and becomes smaller for the consecutive SET pulses. This suggests, that the state drift after SET is dependent on the extent of the state change that preceded the drift. 

Note that the drift occurs in negative direction, i.e., the resistance increases.  Since the measurements pulses are $50$\,mV$\times$\,$200$\,$\mu$s, i.e., positive, any effect the measurement had on the resistance would drive it to lower values. This, apart from the measurement voltage being smaller than one fifth of the average measured threshold voltage, shows that the drift is not caused by the measurement itself, but is a separate effect. {A possible explanation for this drift could be the collapsing of unstable portions of the conductive channel{, a phenomenon observed in~\cite{valov2014redox} too}.} 

\section{Model}
\label{chap:model}

\subsection{Base Model Selection}
\label{sec:mod_selection}
As mentioned before, models that simulate the internal physical respectively electro-chemical effects of memristors, besides being computationally more expensive, are specially designed for a certain memristor type and even certain material pairings of which the modeled memristor is composed.

Behavioral models on the other hand can be used for a family of similarly working memristors to a certain extend. A behavioral model for memristors that is both well-known and well-received by the community is the \gls{vteam} model~\cite{kvatinsky2015vteam}. It models the memristor as a voltage dependent resistance switch that has a continuous, bounded state variable. The change of the state variable depends on the applied voltage, making the model a so-called \textit{voltage-controlled} model, which is reflected in the name of the model, in contrast to its twin brother, the \gls{team} model~\cite{kvatinsky2012team}, which is current-controlled. 
The \gls{vteam} model was used as a starting point for the development of an enhanced memristor model because of the following reasons:
{
(a) It is well-received by the community,
(b) it is one of the most flexible behavioral models,
(c) \gls{vteam} can be implemented\footnote{{VTEAM} was implemented in SPICE at the Institute of Computer Technology, TU Wien by Martin Jungwirth and David Radakovits and refined by Simon Laube~\cite{VTEAM_Spice}.} in pure \gls{spice} without any additional procedural code, for example, C or Verilog-A,
(d) \gls{vteam} was chosen over \gls{team} because in most circuit design situations it is more convenient to use voltage levels than current levels,
(e) In \gls{vteam} modeling of state change is independent from the modeling of voltage-current behavior.
}

{As mentioned above,} an advantage of \gls{vteam} is, that the modeling of state change is in principle independent from the modeling of voltage-current behavior,
as {it} can be seen from the Equations~\eqref{eq:vteam_update_state}~and~\eqref{eq:vteam_update_UI}.

 \begin{samepage}
 
    \begin{equation}
        \label{eq:vteam_update_state}
        \frac{\mathrm{d}w(t)}{\mathrm{d}t} = 
        \begin{cases}
            k_\mathrm{off} \cdot \left( \frac{v(t)}{v_\mathrm{off}} -1 \right)^{\alpha_\mathrm{off}} \cdot f_\mathrm{off}(w) & \text{, } 0<v_\mathrm{off}<v \\
            0 & \text{, } v_{on}<v<v_\mathrm{off} \\
            k_\mathrm{on} \cdot \left( \frac{v(t)}{v_\mathrm{on}} -1 \right)^{\alpha_\mathrm{on}} \cdot f_\mathrm{on}(w) & \text{, } v<v_\mathrm{on}<0
        \end{cases}       
    \end{equation}
    
    \begin{subequations}
    \begin{equation}
        f_\mathrm{off}(w) = e^{-e^{\frac{w(t)-a_\mathrm{off}}{w_\mathrm{c}}}}
    \end{equation}
    \begin{equation}
        f_\mathrm{on}(w) = e^{-e^{\frac{-w(t)-a_\mathrm{on}}{w_\mathrm{c}}}}
    \end{equation}
    \label{eq:vteam_update_window}
    \end{subequations}
    
    \vspace{-3mm}
    \begin{equation}
        v(t)=\left[ R_\mathrm{on}+\frac{R_\mathrm{off}-R_\mathrm{on}}{w_\mathrm{off}-w_\mathrm{on}} \left( w(t)-w_\mathrm{on} \right) \right]\cdot i(t)
    \label{eq:vteam_update_UI}    
    \end{equation}
    
\end{samepage}
\Cref{tab:vteam_parameters} lists all parameters for the \gls{vteam} model. The parameters are briefly explained in the table. Since all parameters are constants during simulations with the \gls{vteam} model -- only the state $w(t)$ and the voltage, $v(t)$, and consequently, the current, $i(t)$, are variables -- the model is not capable of simulating any variation in \gls{hrs} and \gls{lrs} resistance or state dynamics. {In order to do so},  this work introduces additions to this model in Sections~\ref{sec:mod_variation}~and~\ref{sec:mod_leakage} to enhance the model in terms of adding those capabilities.
\begin{table}
    \centering
    \caption{\gls{vteam} parameters }
    \begin{tabular}{|c|l|}
    \hline
        Parameter       & Explanation \\
    \hline
        $v_{off}$       & Positive threshold voltage.\\
        $v_{on}$        & Negative threshold voltage.\\
        $k_{off}$       & State change rate during SET.\\
        $k_{on}$        & State change rate during RESET.\\
        $\alpha_{off}$  & Degree of (non-)linearity for SET.\\
        $\alpha_{on}$   & Degree of (non-)linearity for RESET.\\
        $a_{off}$       & Window boundary for SET.\\
        $a_{on}$        & Window boundary for RESET.\\
        $w_c$           & Shaping parameter for window function.\\
        $w_{off}$       & Value of state variable at \gls{hrs}.\\
        $w_{on}$        & Value of state variable at \gls{lrs}.\\
        \Roff{}       & Resistance at \gls{hrs}.\\
        \Ron{}        & Resistance at \gls{lrs}.\\
    \hline
    \end{tabular}
    \label{tab:vteam_parameters}
\end{table}

\subsection{Variation Model}
\label{sec:mod_variation}

Instead of setting the model parameters to constant values, the \gls{spice} functions \texttt{.gauss()} and \texttt{.flat()} are used to model normal {and uniform} distribution of parameters{, respectively}.
The functions expect a mean value and distribution interval (standard deviation in the case of normal distribution) and automatically pick 
respective parameters randomly 
using the chosen distribution.
This enables the simulation of both cycle-to-cycle and device-to-device variations in memristors. This method is used to model the variations in \gls{hrs} and \gls{lrs} resistance, resistance change dynamics and threshold voltages.

\subsection{Leakage Model}
\label{sec:mod_leakage}
In order to simulate leakage, i.e., the drift of the memristor state in the absence of any stimulus, the model was extended by an additional influence on the state variable, $w(t)$. As {it} was shown in \Cref{sec:exp_leakage}, the drift depends on the extend of state change that happened during the SET process and fades out after a certain amount of time. Therefore, the state derivative, $\frac{\mathrm{d}w(t)}{\mathrm{d}t}$ in \Cref{eq:vteam_update_state}, was extended by a term, that represents a decaying integral of state change. Equations~(\ref{eq:state})~and~(\ref{eq:leakage}) describe the enhanced model of state dynamics. The extension to the \gls{vteam} model is given in blue. \Cref{eq:leakage} in particular introduces a new term, which represents a capacity $\Theta(t)$, which integrates the state change and has a constant drain via the introduced model parameter $\tau_\mathrm{L}$. The introduced parameters $\theta_\mathrm{off}$~and~$\theta_\mathrm{on}$ describe the influence of state change on the initial drift rate during SET and RESET process, respectively. Note that the model in principle is capable of modeling a state drift after RESET, however by setting $\theta_\mathrm{on}=0$ this feature is disabled. 
As {it} can be seen in \Cref{eq:leakage}, $\Theta(t)$ depends on the state change of the memristor that is 
{calculated}
the same way as  
the \gls{vteam} model, except for the additional factors $\theta_\mathrm{off}$~and~$\theta_\mathrm{on}$.
The modeling of voltage-current behavior is unaltered and can be seen in \Cref{eq:vteam_update_UI}.


\begin{table*}
\caption{New Model Equations}
\centering

\begin{minipage}[]{\textwidth}

\begin{subequations}
    \begin{align}
        \label{eq:state}
        \frac{\mathrm{d}w(t)}{\mathrm{d}t} =
         \begin{cases}
            k_\mathrm{off} \cdot \left( \frac{v(t)}{v_\mathrm{off}} -1 \right)^{\alpha_\mathrm{off}} \cdot f_\mathrm{off}(w) & \text{, } 0<v_\mathrm{off}<v(t) \\
            \textcolor{blue}{ \Theta(t) } & \text{, } v_\mathrm{on}<v(t)<v_\mathrm{off} \\
            k_\mathrm{on} \cdot \left( \frac{v(t)}{v_\mathrm{on}} -1 \right)^{\alpha_\mathrm{on}} \cdot f_\mathrm{on}(w) & \text{, } v(t)<v_\mathrm{on}<0
        \end{cases}
    \end{align}
    \begin{align}
         \label{eq:leakage}
        \textcolor{blue}{
        \frac{\mathrm{d}\Theta(t)}{dt} = -\frac{\Theta(t)}{\tau_\mathrm{L}} +
        \begin{cases}
            \theta_\mathrm{off} \cdot k_\mathrm{off} \cdot \left( \frac{v(t)}{v_\mathrm{off}} -1 \right)^{\alpha_\mathrm{off}} \cdot f_\mathrm{off}(w) & \text{, } 0<v_\mathrm{off}<v(t)\text{, } \theta_\mathrm{off}>0 \\
            0 & \text{, } v_\mathrm{on}<v(t)<v_\mathrm{off} \\
            \theta_\mathrm{on} \cdot k_\mathrm{on} \cdot \left( \frac{v(t)}{v_\mathrm{on}} -1 \right)^{\alpha_\mathrm{on}} \cdot f_\mathrm{on}(w) & \text{, } v(t)<v_\mathrm{on}<0\text{, } \theta_\mathrm{on}<0
        \end{cases}
        }
    \end{align}
    \label{eq:update_state}
    \end{subequations}
\end{minipage}
\end{table*}
    

\begin{table} [!b]
    \centering
    \caption[New parameters introduced to the model]{New parameters introduced to the model.}
    \vspace{3mm}
    \begin{tabular}{|c|c|}
    \hline
    Parameter & Explanation \\
    \hline
        $\theta_\mathrm{off}$ & Influence of SET process state change on leakage \\
        $\theta_\mathrm{on}$ & Influence of RESET process state change on leakage \\
        $\tau_\mathrm{L}$ & Time constant of leakage fade-out \\
    \hline
    \end{tabular}
    \label{tab:new_parameters}
\end{table}{}

\subsection{Fitting}
\label{sec:mod_fitting}
This section explains the process of model fitting in four parts: Resistance in \gls{hrs} and \gls{lrs}, resistance change rate, threshold voltages and the observed leakage effect. The individual parameter values that where determined during the fitting process are collectively given in \Cref{subsec:parameters}.

\subsubsection{{HRS and LRS} Resistance}
\label{subsec:hrs_lrs}
As {it was} shown before in \Cref{sec:exp_RON_ROFF}, the resistance in \gls{hrs} and \gls{lrs} can be modeled via a Gaussian or normal distribution with the parameters from \Cref{tab:model_values}.

\begin{figure}[b!]
    \centering
    \includegraphics[width=\linewidth]{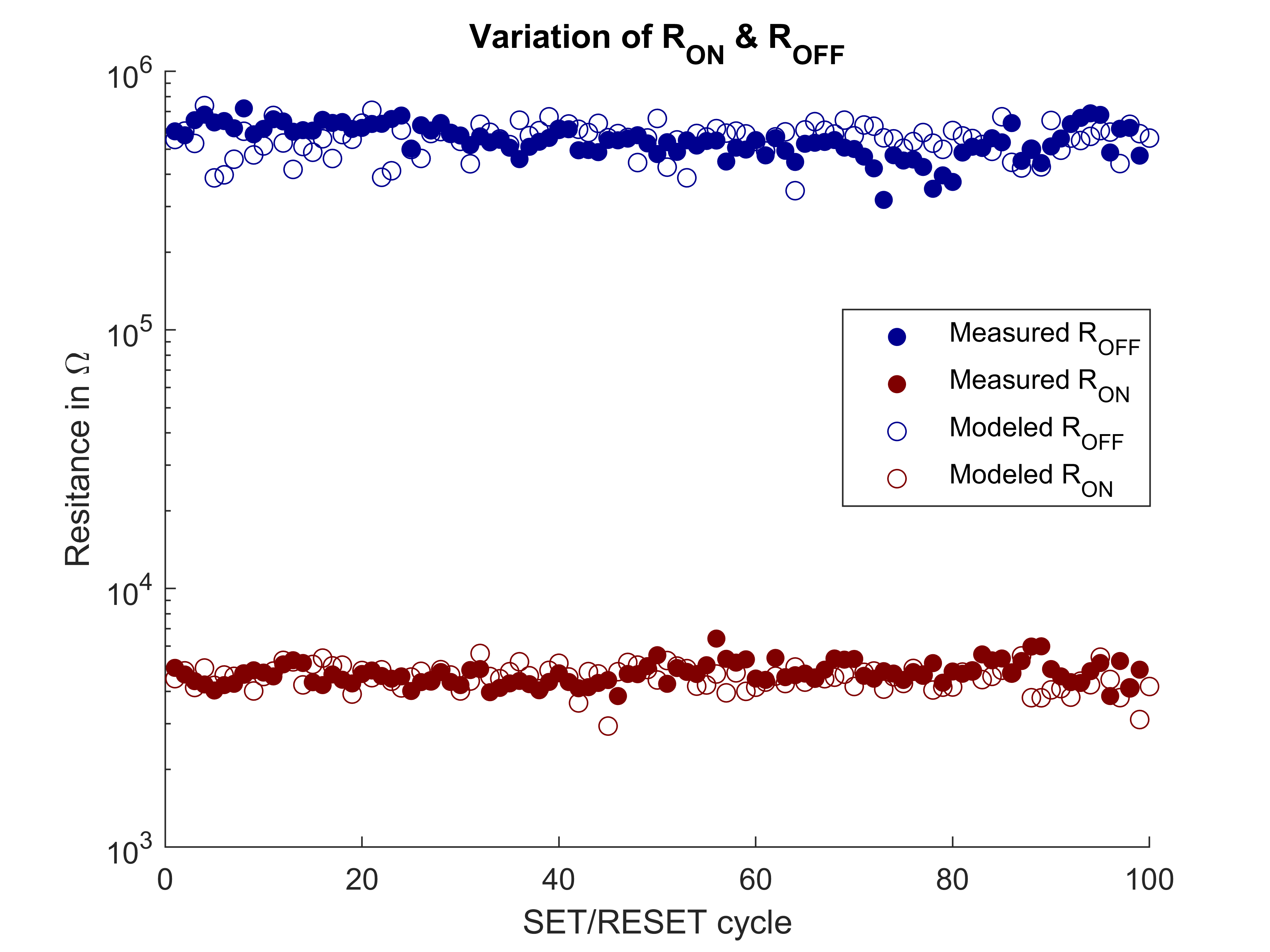}
    \caption[Example of model versus measured \gls{hrs} and \gls{lrs} resistances]{Example of model vs. measured \gls{hrs} and \gls{lrs} resistances.}
    \label{fig:res_var_model_data}
\end{figure}

\Cref{fig:res_var_model_data} shows an example of the measured \gls{hrs} and \gls{lrs} resistance versus the simulated variation.

\subsubsection{Resistance Change Dynamics}
\label{subsec:change_rate}
\begin{figure}
    \centering
    \includegraphics[width=\linewidth]{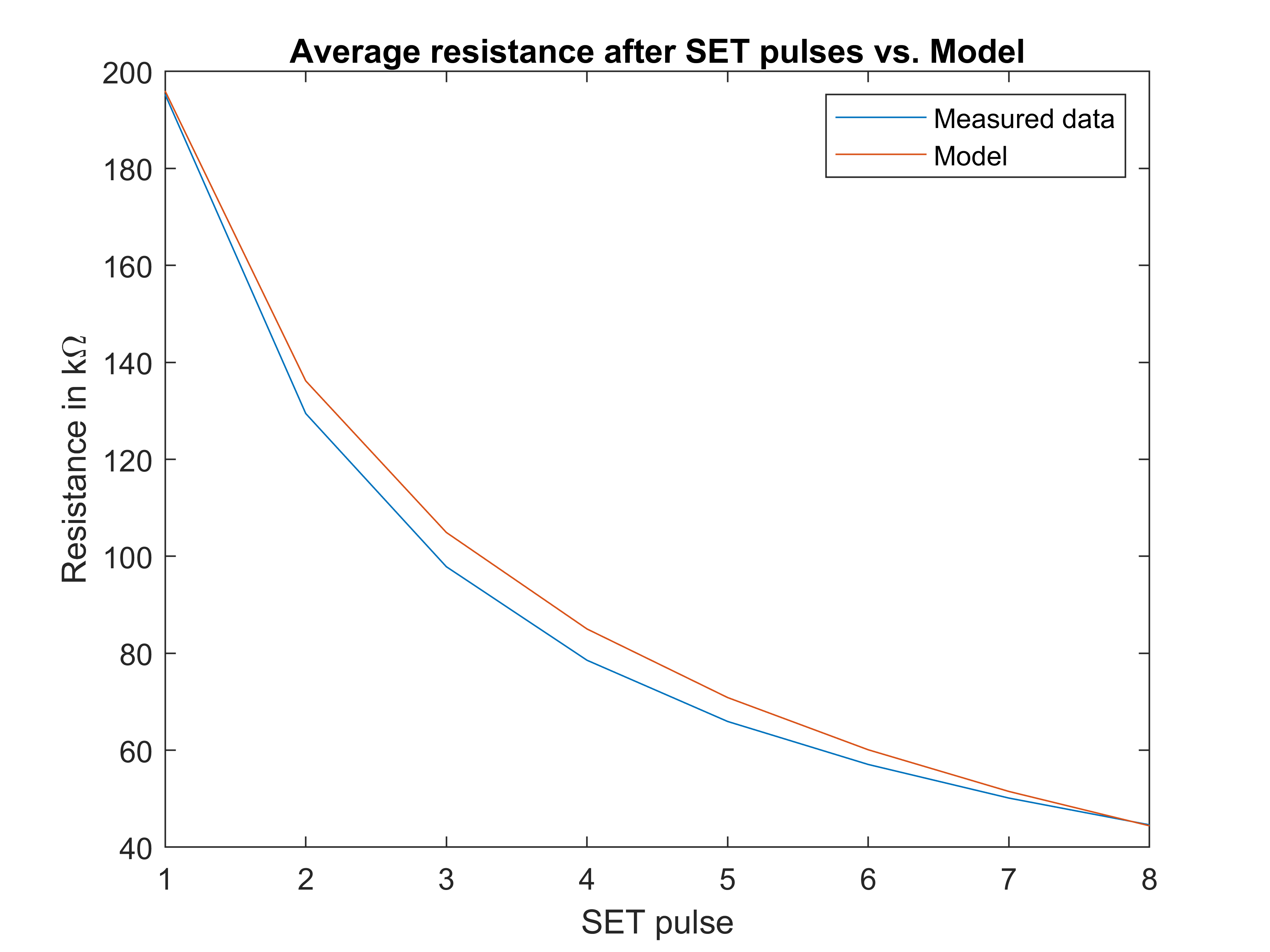}
    \caption[Measured average resistance change during SET process versus model]{Measured resistance change during SET process (average of $100$ measurements) versus model. 
    }
    \label{fig:res_change_model}
\end{figure}

The rate of resistance change is influenced by several parameters in the model for both SET and RESET process, respectively. These parameters are namely $k_\mathrm{off}$, $k_\mathrm{on}$ $a_\mathrm{off}$, $a_\mathrm{on}$ and $w_\mathrm{c}$. 
While $k_\mathrm{off}$ and $k_\mathrm{on}$ {scale} 
the state change linearly,
$a_\mathrm{off}$, $a_\mathrm{on}$ and $w_c$ are used to influence the shape of the state change, i.e., to dampen the state change in proximity of state boundaries.
\Cref{fig:res_change_model} shows the fitting of the shape parameters $a_\mathrm{off}$, $a_\mathrm{on}$ and $w_\mathrm{c}$ and the average rate of state change during SET, which is $k_\mathrm{off}=780$\,$\mu$m/s.

\begin{figure}
    \centering
    \includegraphics[width=0.8\linewidth]{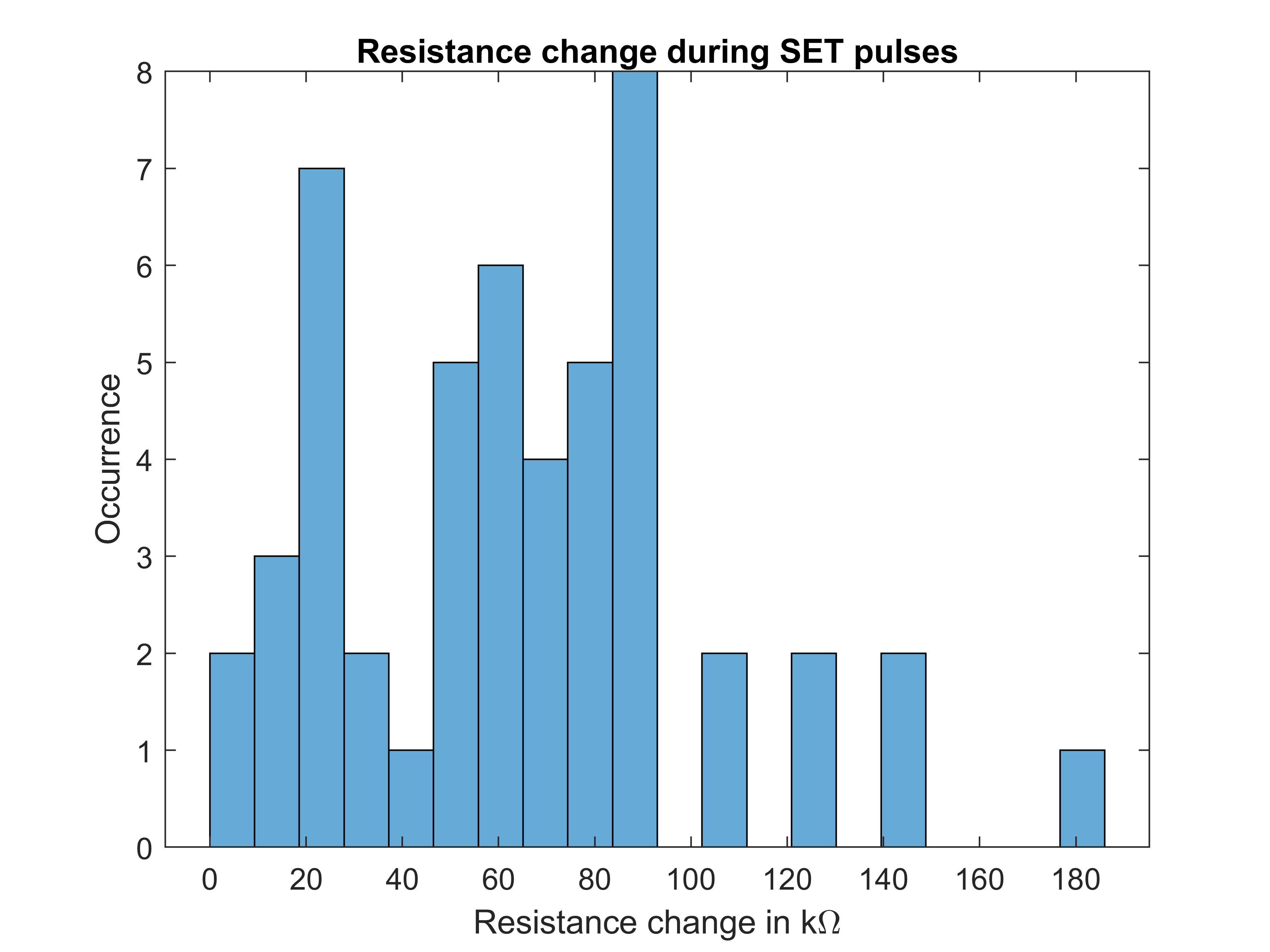}
    \caption[Histogram of measured resistance change during SET]{Histogram of measured resistance change during SET, {measured as difference between third and first measurement pulse.} Note, that resistance change from \gls{hrs} to \gls{lrs} is considered positive in this figure.
    }
    \label{fig:res_change_hist}
\end{figure}

The histogram of the recorded resistance change is given in \Cref{fig:res_change_hist}. The resistance change shown in this figure was calculated as the difference between the resistance after the first of the eight applied SET pulses and the resistance after the third pulse, 
{since in the fourth and fifth pulse, the change is not linear anymore.}
Since the distribution of resistance change, that was measured and shown in \Cref{fig:res_change_hist}, does not 
assimilate well-known distribution models, a different approach   
(compared to that of \gls{hrs} and \gls{lrs} resistance) was chosen. As {it} can be seen in the figure, $75$\,\% of the measured values lie in the interval from $18.4$\,k$\Omega$ to $93$\,k$\Omega$ with a steep decrease of measured values above and below this interval. {Therefore,} 
this interval was chosen for {a uniform distribution} of the resistance change variation during SET. 
With an average measured resistance change of $55.8$\,k$\Omega$ the variation calculates to $\pm 67$\,\%, which translates into a $k_\mathrm{off}$ variation of $\pm 522.6$\,$\mu$m/s ($3\sigma)$.

\begin{figure}
    \centering
    \includegraphics[width=\linewidth]{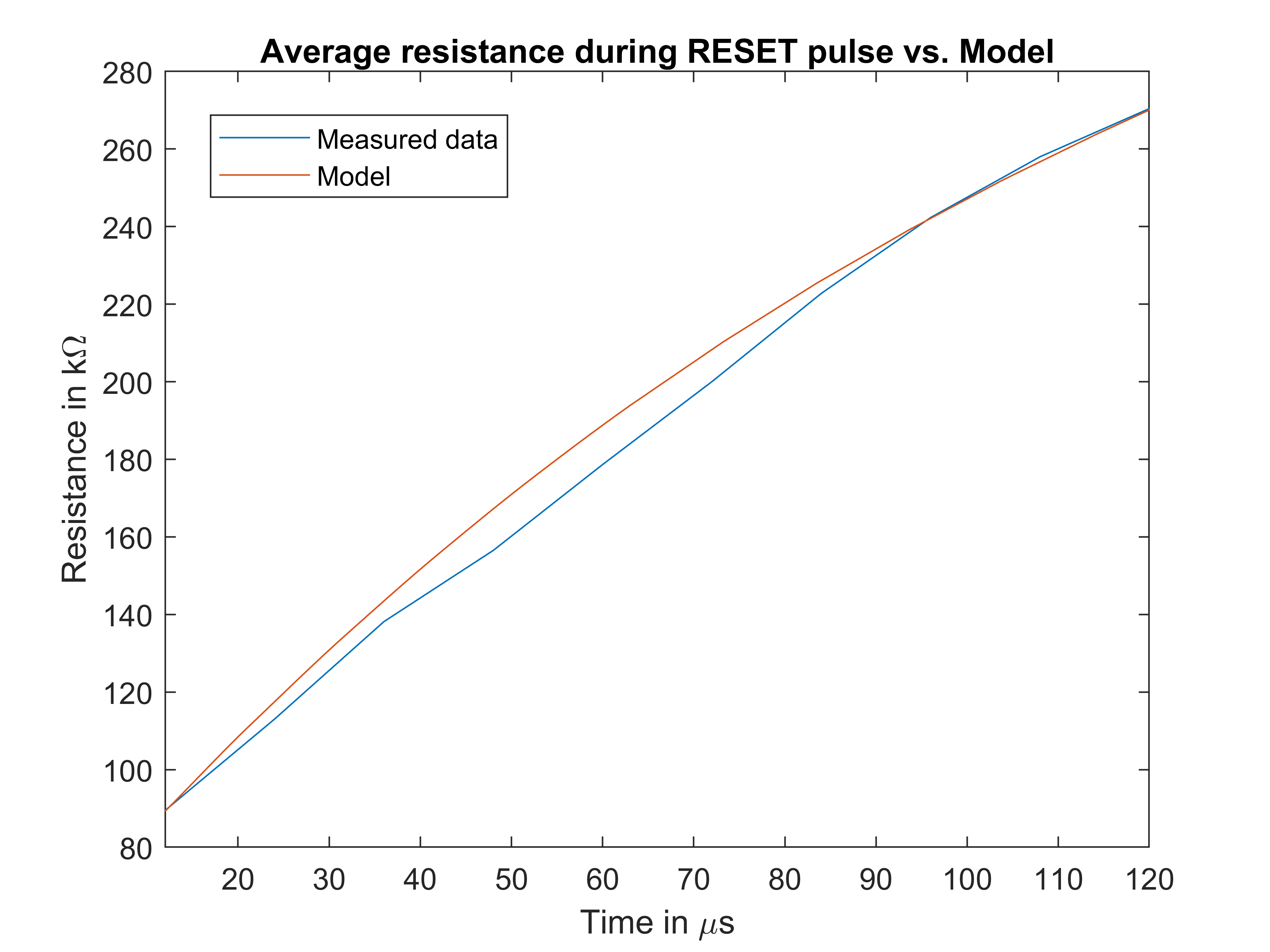}
    \caption[Average measured resistance during RESET]{Resistance measured during RESET pulse (average of $100$ measurements) versus model.
    }
    \label{fig:RESET_change_model}
\end{figure}

\begin{figure}
    \centering
    \includegraphics[width=0.8\linewidth]{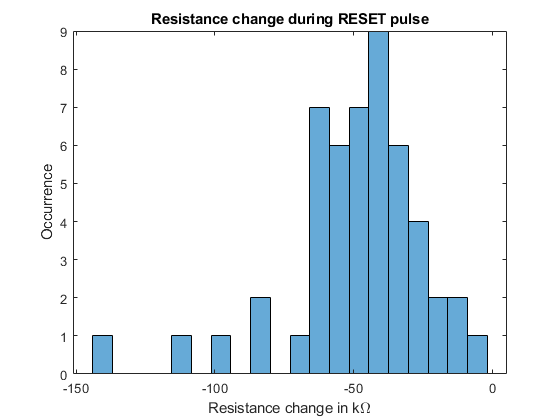}
    \caption[Histogram of measured resistance change during RESET]{Histogram of measured resistance change during RESET. Note, that resistance change from \gls{lrs} to \gls{hrs} is considered negative in this figure.}
    \label{fig:RESET_hist}
\end{figure}

\Cref{fig:RESET_change_model} shows the fitting of the average rate of state change during RESET, which 
matches $k_\mathrm{on}=-4.67$\,$\mu$m/s. The same shaping parameters are used for both SET and RESET process. The variation of the state change during RESET was determined in the same manner as described above for the SET process. \Cref{fig:RESET_hist} shows the histogram of the measured data. More than $75$\,\% of the measurements lie between $-65.9$\,k$\Omega$ and $-23.3$\,k$\Omega$.
With an average measured resistance change of $-44.6$\,k$\Omega$ the variation calculates to $\pm 48$\,\%, which translates into a variation of $k_\mathrm{on}$ of $\pm 2.24$\,$\mu$m/s ($3\sigma$).

\Cref{tab:model_values} gives the statistical parameters for the resistance change variation model.

\subsubsection{Threshold Voltage}
\label{subsec:threshold}
The actual threshold behavior of the model is influenced not only by the $v_\mathrm{off}$ and $v_\mathrm{on}$ parameters, but also by the exponents $\alpha_\mathrm{off}$ and $\alpha_\mathrm{on}$, which influence the sharpness of the threshold. 
The threshold voltages
were fitted using the recorded hystereses to mimic the observed behavior.

The values for the exponents $\alpha_\mathrm{on}$ and $\alpha_\mathrm{off}$ were taken from the original \gls{vteam} publication\cite{kvatinsky2015vteam}.
The average threshold voltages, as reported in \Cref{sec:exp_threshold}, are $v_\mathrm{off}=370.2$\,mV and $v_\mathrm{on}=-373.8$\,mV.
The largest deviation from the average was $30.57$\,\% for $v_\mathrm{off}$ and $33.01$\,\% for $v_\mathrm{on}$. These values were used as a variation interval ($3\sigma$) for
{a variation model with uniform distribution,}
{since the actual distribution could not be extracted based on the data set at hand. As explained in \Cref{subsec:change_rate} this approach is suitable to give a worst case estimation of threshold voltage variation.}
\Cref{tab:model_values} gives the parameters for the distribution.

\subsubsection{Leakage}
\label{subsec:leakage}
The parameters that influence the modeling of the leakage effect are $\theta_\mathrm{off}$, $\theta_\mathrm{on}$ and $\tau_L$. As mentioned in \Cref{sec:exp_leakage}, no leakage effect was observed after RESET. Therefore, $\theta_\mathrm{on}$ can be set to zero. $\theta_\mathrm{off}$ and $\tau_L$ were set to mimic the observed behavior in the experiments.
In order to do that, since the leakage model describes an exponential drift which at $t=5\tau$ differs for less than $0.1$\,\% from its value for $t \rightarrow \infty$, $\tau_L$ is set to a fifth of the observed duration of state change due to leakage.
{\Cref{fig:retention_1}} shows an example of the fitted model versus the data measured during a leakage measurement. \Cref{fig:retention_1_series} shows another type of experiment, in which a series of SET pulses was applied.
As {it} can be seen from the figure, the model is capable of reproducing the observed behavior. The average values observed during all leakage experiments are $\theta_\mathrm{off}=0.0173$\,s$^{-1}$, $\tau_L=10.3$\,s and $\theta_\mathrm{on}=0$, since no leakage effect was observed after RESET.

\subsubsection{Summary}
\label{subsec:parameters}
The developed memristor model, that incorporates both the variation of parameters and the leakage effect, is based on the \gls{vteam} model, as is given in Equations~\eqref{eq:vteam_update_state},~\eqref{eq:vteam_update_window}~and~\eqref{eq:vteam_update_UI}. 
In the enhanced model, the memristor is also modeled as a linearly state-dependent, linear voltage-current relationship, cf.~\Cref{eq:vteam_update_UI}.

The state dynamics are non-linearly dependent on the voltage applied to the memristor, cf.~\Cref{eq:state}.

In this, $f_\mathrm{off}$ and $f_\mathrm{on}$ denote window functions, which are used to {dampen} the state change in the proximity of the state limits, cf.~\Cref{eq:vteam_update_window}.

\Cref{eq:state} holds an additional term over the original \gls{vteam} state equations, which is $\Theta(t)$. This term denotes the influence of leakage on the memristor state. $\Theta(t)$ is calculated as a decaying state drift, which is dependent on the previous change of memristor state and the current memristor state, cf.~\Cref{eq:leakage}.

The modeling of parameter variations is achieved by a dynamic calculation of core model parameters, namely $R_\mathrm{ON}$, $R_\mathrm{OFF}$, $k_\mathrm{ON}$, $k_\mathrm{OFF}$, $v_\mathrm{ON}$ and $v_\mathrm{OFF}$, using distribution functions.
\Cref{tab:model_values} gives the final values for all model parameters. Note that the variation of the resistance change rates $k_\mathrm{on}$~and~$k_\mathrm{off}$ and the variation of threshold voltages are a third of the values given in Sections~\ref{subsec:change_rate}~and~\ref{subsec:threshold}, since both \texttt{.gauss()} and \texttt{.flat()} expect the standard deviation $\sigma$ as a parameter, whereas in Sections~\ref{subsec:change_rate}~and~\ref{subsec:threshold} the overall variation interval ($3\sigma$) is given.
{The full \gls{spice} model code is given in the appendix.}

\begin{table}[b!]
    \centering
    \caption[Values for all model parameters]{Values for all model parameters.}
    \begin{tabular}{|c|c|c|c|c|}
    \hline
        Parameter     & \multicolumn{3}{c|}{ Value } \\
    \hline
        Distribution  & Mean & Standard deviation & Type \\
    \hline
        \Roff{}       & $545.54$\,k$\Omega$ & $ 77.095$\,k$\Omega$    & gaussian \\
        \Ron{}        & $4.92$\,k$\Omega$   & $ 858.8$\,$\Omega$      & gaussian \\
        $v_\mathrm{off}$       & $ 370.2$\,mV        & $ 37.7$\,mV     & uniform    \\
        $v_\mathrm{on}$        & $ -373.8$\,mV       & $ 41.1$\,mV     & uniform    \\
        $k_\mathrm{off}$       & $ 780$\,$\mu$m/s    & $ 174.2$\,$\mu$m/s  & uniform\\
        $k_\mathrm{on}$        & $ -4.67$\,$\mu$m/s  & $ 0.747$\,$\mu$m/s  & uniform\\
    \hline
        $\alpha_\mathrm{off}$  & \multicolumn{3}{c|}{ $3$ }                \\
        $\alpha_\mathrm{on}$   & \multicolumn{3}{c|}{ $3$ }                \\
        $a_\mathrm{off}$       & \multicolumn{3}{c|}{ $1.3$\,nm }          \\
        $a_\mathrm{on}$        & \multicolumn{3}{c|}{ $1.8$\,nm }          \\
        $w_c$                  & \multicolumn{3}{c|}{ $980$\,pm }          \\
        $w_\mathrm{off}$       & \multicolumn{3}{c|}{ $3$\,nm }            \\
        $w_\mathrm{on}$        & \multicolumn{3}{c|}{ $0$\,nm }            \\
        $\theta_\mathrm{off}$  & \multicolumn{3}{c|}{ $0.0173$\,s$^{-1}$ } \\
        $\theta_\mathrm{on}$   & \multicolumn{3}{c|}{ $0$\,s$^{-1}$ }      \\
        $\tau_L$               & \multicolumn{3}{c|}{ $10.3$\,s }          \\
    \hline
    \end{tabular}
    \label{tab:model_values}
\end{table}

\section{Discussion}
\label{sec:mod_discussion}
The model presented in this work is capable of simulating the leakage effect, which is a novelty among behavioral models. The simulation {inaccuracy}
of the leakage effect is $\leq 13.4$\,\% for all points in all 20 tested cases. The average {of absolute values of relative} deviation
of the model from the data is $1.1$\,\%.
{This average deviation was calculated as}
\begin{equation}
\label{eq:average_deviation}
    \Delta=\frac{1}{N}\sum_{i=1}^N \frac{|\Tilde{x_i}-x_i|}{x_i}
\end{equation}
{where $\Tilde{x_i}$ denotes the modeled value and $x_i$ denotes actual, measured value.}
It needs to be stated, that similar to all other parameters of the memristor, both the time constant of the leakage 
decay, as well as the initial leakage strength varied over the course of the investigation. However, 
{the effect of those variations is small, i.e., the accuracy of the model was not affected dramatically and the variations were not included in the model to simplify the implementation.}

The simulation {inaccuracy} of the state change rate is $\leq 8.2$\,\% for all points and $4.6$\,\% on average over the $100$ tested cases.
{The average deviation was again calculated according to \Cref{eq:average_deviation}.}

The identification and fitting of threshold voltages is specially hard to manage. The optimal experiment to identify a memristors threshold voltage, e.g., the positive threshold voltage, would consist of the application of a pulsed stimulus, in which the pulses have constant duration but increasing amplitude. 
The threshold voltage 
{is defined as the voltage}
{that causes the resistance}
change {to exceed} a certain defined value. Keep in mind, that the measurement of resistance would need to be carried out with measurement pulses in between the stimuli, to ensure suppression of any distortion due to non-linear voltage-resistance behavior. However, since the momentary resistance of a memristor depends on its past, the experiment would need to be repeated with only the stimulus that was just too small and the stimulus that caused a significant state change, to suppress the effect of the preceding stimuli. Since most probably the threshold voltage might vary during these measurements, a successful determination of the threshold voltage is still not guaranteed.

The model as it is presented and implemented in SPICE is capable of modeling the variation of the mentioned parameters in both cycle-to-cycle as well as device-to-device fashion. However, it is not possible to model parameter variations during simulation runs, as the parameters are newly chosen for each cycle but remain constant during a cycle. Simulation of device-to-device variations is achieved via separate selection of parameters for each individual device. The identification of device-to-device variations, however, is out of the scope of this work.

\section{Case Studies}
\label{chap:casestudy}
Several memristive logics have been presented in the literature, some of the prominent ones being \gls{imply}\cite{borghetti2010memristive}, \gls{magic}\cite{Kvatinsky2014magic}, \gls{felix}\cite{Gupta2018} and \gls{tmsl}\cite{huang2016reconfigurable}. All these logics are \textit{stateful}, i.e., the input and output logic values are represented by the resistance of a memristor and all of them function based on the application of certain, predefined voltages to calculate the desired state of the output memristor. In this chapter each of the four logics is used to design a logic gate under the specifications of the respective publication. In order to show the effect of parameter variations and leakage on the designed gates, each individual gate is designed with the nominal parameter values, i.e., the mean values, and without consideration of leakage, as it is done in the publications. Probabilities of correct output calculation are then determined through simulation of the memristive logic gates with the model that incorporates both parameter variations and leakage.

Note, that for all the investigated logics, there are certain degrees of freedom, e.g., execution time or the selection of certain voltages from an allowed interval, which result in different trade-offs for speed or different aspects of robustness. However, it is out of the scope of this work, to find an optimal gate design. 
Here, the design rules given in the publications are followed, i.e., an allowed parameter set is chosen and tested for functional correctness for all input combinations for each logic, and the impact of parameter variations and leakage is studied with the obtained gate designs in order to stress the importance of incorporating those non-idealities into memristor models. 
All investigations are conducted based on a logic threshold of $0.5$, i.e., 
$0 \leq s < 0.5\textnormal{ }\widehat{=}$\,`0' and $0.5 \leq s \leq 1\textnormal{ }\widehat{=}$\,`1',
where $s$ denotes the state of the memristor,
was chosen as a mapping of memristor state to logical values. Overall correctness of logical operation was calculated as the overall probability of correctly processed outputs, i.e, $P_\mathrm{correct}=\frac{1}{4}P_\mathrm{00}+\frac{1}{4}P_\mathrm{01}+\frac{1}{4}P_\mathrm{10}+\frac{1}{4}P_\mathrm{11}$, where $P_\mathrm{correct}$ is the overall correctness, and $P_\mathrm{00}$ to $P_\mathrm{11}$ are the probabilities of the individual input combinations to yield correct outputs.

{Furthermore, the effect of leakage on the reliability of the gates under investigation was studied. In order to do so, a stable time was defined, which gives the time span after the logic operation within which the result of the logic operation remains unaltered by leakage, i.e., {after the stable time the result becomes incorrect because the leakage caused the memristor state to cross the chosen logic threshold and a so-called bit flip occurs.}}

\subsection{IMPLY}
\label{sec:case_IMPLY}
Material Implication Logic (\gls{imply}), as the name suggests, implements a material implication function.
{The $q$ memristor, which holds both the second operand before, as well as the output after the logic operation, only changes its state in the case $0\rightarrow 0$, where the stored bit goes from `0' to `1'.}
That means that since in \gls{imply} the common mapping of$\textnormal{ }$ `0' $\widehat{=}$ {HRS} and `1' $\widehat{=}$ {LRS} is used, the $q$ memristor has to be turned on, i.e., SET, in Case 1, where both memristors are in {HRS} prior to the logic operation and must not be turned on, when only $p$ is in {LRS} as in Case 3. \Cref{fig:IMPLY_circ} shows the circuit of a basic \gls{imply} gate with $V_\mathrm{set} > v_\mathrm{off}$, $V_\mathrm{set} > V_\mathrm{cond}$ 
and $R_\mathrm{on} < R_\mathrm{G} < R_\mathrm{off}$. Additional design constraints are given in~\cite{kvatinsky2011memristor}. \Cref{tab:logic_param} gives the logic parameters used for the simulation of \gls{imply} gates.

\begin{figure*}[!ht]
    \begin{minipage}[l]{0.6\columnwidth}
        \centering


\tikzset{varr/.style={-stealth}}
\begin{circuitikz}[scale=.68,transform shape]
					
					\draw(0,-3.5) to[short,-*] (1.5,-3.5)
					     (3,-3.5) to[short] (1.5,-3.5)
					     (1.5,-3.5) to[short] (1.5,-4)
					     (1.5,-4) to[R,l_=$R_\mathrm{G}$,font=\Large] (1.5,-5.5) node[rground]{};
					\draw(0,-3.5) to[Mr,-o,l^=$P$,font=\Large] (0,-1) node[above]{$V_\mathrm{cond}$}
					(3,-3.5) to[Mr,-o,l^=$Q$,font=\Large](3,-1) node[above]{$V_\mathrm{set}$};	
				\end{circuitikz}

        \caption[Circuit of basic {IMPLY} gate]{Circuit of basic {IMPLY} gate. $Q$ holds the result of $p\rightarrow q$ after the operation.}
        \label{fig:IMPLY_circ}
    \end{minipage}
    \hfill{}
    \begin{minipage}[r]{0.6\columnwidth}
        \centering

\def\xlen{4mm}


\begin{circuitikz}[scale=.68,transform shape]
	\draw (0,2) to [Mr,l_=$out$] (0,0) node[rground]{}
	(0,2) to[short,*-](-1,2) to[Mr,l^=$in_\mathrm{1}$] (-1,4) to[short,-*] (0,4)
	(0,2) to[short,*-](1,2) to[Mr,l^=$in_\mathrm{2}$] (1,4) to[short,-*] (0,4)
	(0,4) to[short,-o] (0,5)node[above]{\large $V_0$};
\end{circuitikz}

        \caption[Circuit of 2-bit \gls{magic} gate]{Circuit of 2-bit \gls{magic} gate. $in_\mathrm{1}$ and $in_\mathrm{2}$ hold the inputs, $out$ is initialized to \gls{lrs} and holds the output after logic operation.}
        \label{fig:MAGIC_circ}
    \end{minipage}
    \hfill{}
    \begin{minipage}[r]{0.6\columnwidth}
        \centering

\def\xlen{4mm}


\begin{circuitikz}[scale=.68,transform shape]
	
	\draw (0,-2) to [Mr,l_=$out$] (0,0)
	(0,0) to[short,-o] (0,0.5) node[above]{\large $V_0$}
	(0,-2) to[short,*-](-1,-2) to[Mr,l^=$in_\mathrm{1}$] (-1,-4) to[short,-*] (0,-4)
	(0,-2) to[short,*-](1,-2) to[Mr,l^=$in_\mathrm{2}$] (1,-4) to[short,-*] (0,-4)
	(0,-4) to[short,-] (0,-4.5)node[rground]{};
\end{circuitikz}

        \caption[Circuit of 2-bit \gls{felix} OR-gate]{Circuit of 2-bit \gls{felix} OR-gate. $in_\mathrm{1}$ and $in_\mathrm{2}$ hold the inputs, $out$ is initialized to \gls{hrs} and holds the output after logic operation.}
        \label{fig:FELIX_circ}
    \end{minipage}
\end{figure*}

\begin{figure}[t!]
    \centering
    \includegraphics[width=0.8\linewidth]{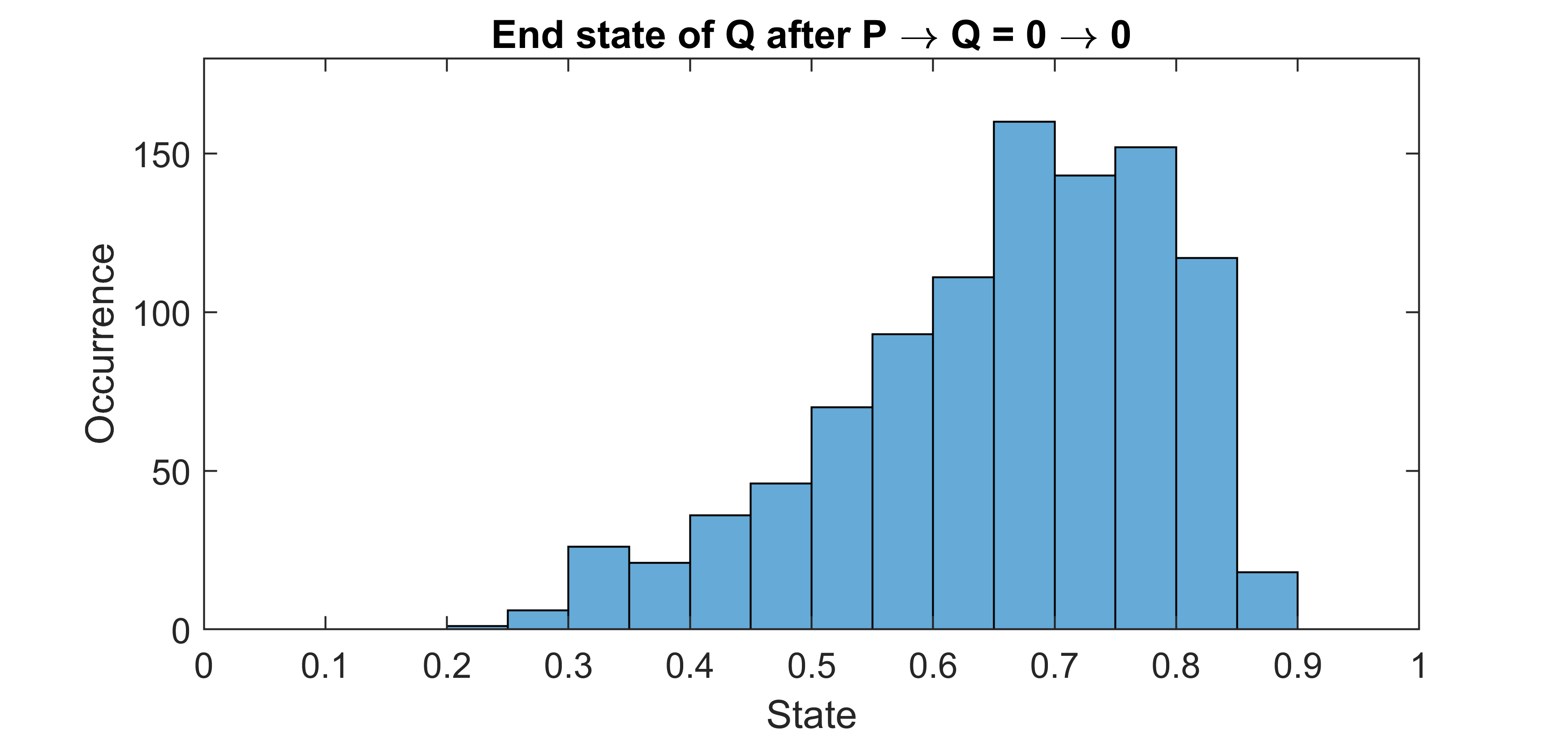}
    \caption[\gls{imply}: State distribution of output memristor]{\gls{imply}: State distribution of output memristor after logic operation for input `00'.}
    \label{fig:IMPLY_00_vari}
\end{figure}

\begin{figure}[t!]
    \centering
    \includegraphics[width=\linewidth]{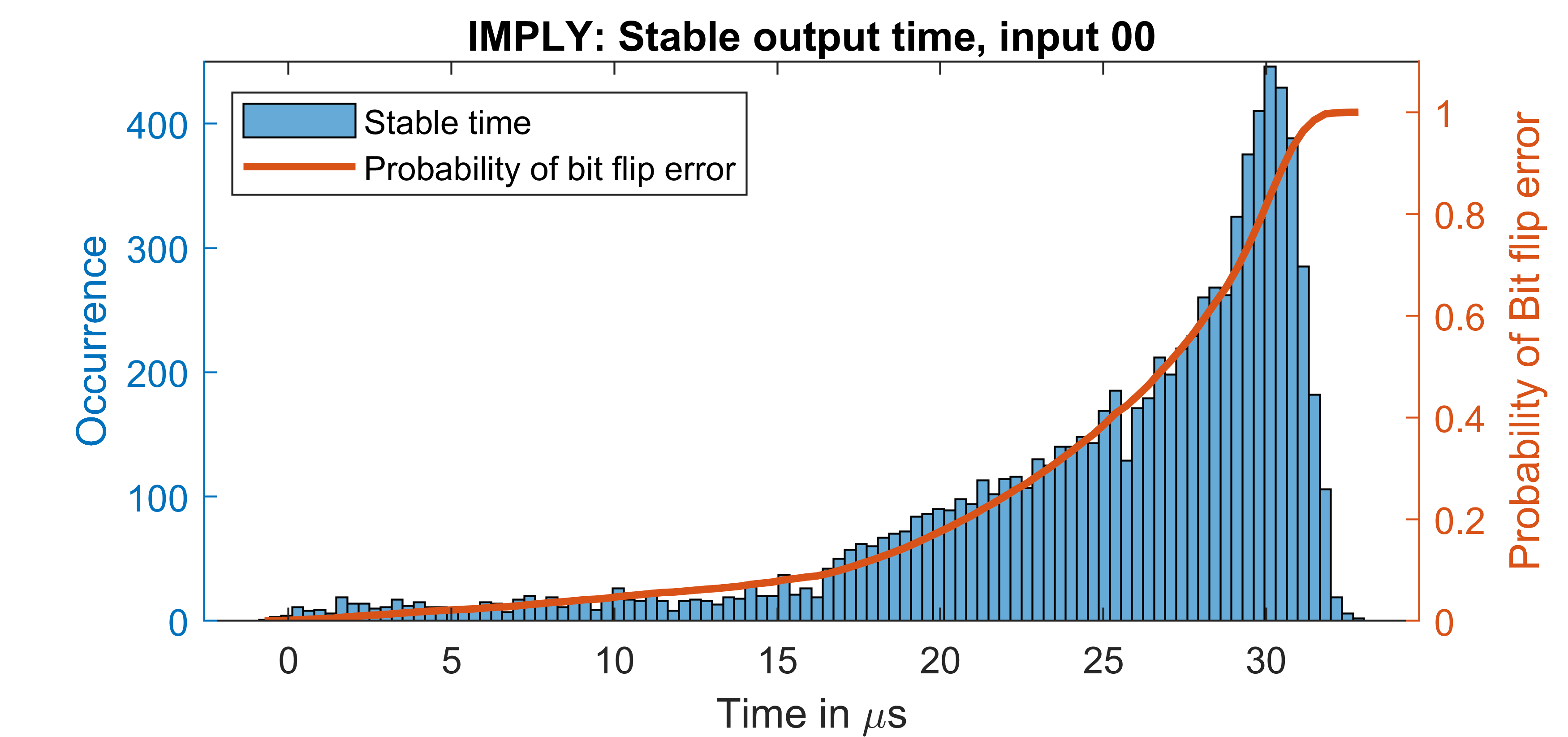}
    \caption[\gls{imply}: Distribution of stable output time for input `00']{\gls{imply}: Distribution of stable output time for input `00'. The orange line gives the accumulated portion of incorrect outputs over time.
    }
    \label{fig:IMPLY_00_leak}
\end{figure}

In \gls{imply} only Case 1, i.e., the input combination `00' is affected by parameter variations and leakage. Cases 2, 3 and 4 are correct in $100$\,\% of the simulated cases. This is due to the inherent function of \gls{imply}, in which only in Case 1 of the truth table a switching occurs. \Cref{fig:IMPLY_00_vari} shows the state histogram of the output memristor $Q$ after logic operation. $86.4$\,\% of the simulations resulted in the correct output, i.e., `1'. Considering the correctness of input combinations `01',`10' and `11', which were always correct, as mentioned above, the overall probability of correct behavior calculates to $96.6$\,\%. As for the parameter variation, also the leakage only affects Case 1 of the \gls{imply} gate truth table. \Cref{fig:IMPLY_00_leak} gives the histogram of the time the output was stable during the simulations. The orange line in the figure gives the accumulated relative portion of all outputs being distorted by leakage. Using this figure, a designer can determine the maximum stable time for a certain probability of a correct output, i.e., that the output was not distorted by leakage. For example, if $>90$\,\% of the calculated outputs shall remain correct, the stable time is $16.9$\,$\mu$s, for $>99$\,\%. the stable time is $1.98$\,$\mu$s.


\subsection{MAGIC}
\label{sec:case_MAGIC}
{Memristor-Aided loGIC} (\gls{magic}) implements a 2-bit {NOR} gate using three memristors. As opposed to \gls{imply}, a separate output memristor is used, as {it} can be seen in \Cref{fig:MAGIC_circ}. This output memristor is initialized to \gls{lrs}, i.e, `1' prior to logic operation. \Cref{tab:logic_param} gives the logic parameters used for the simulation of \gls{magic} gates.

\begin{figure}[t!]
    \centering
    \includegraphics[width=0.8\linewidth]{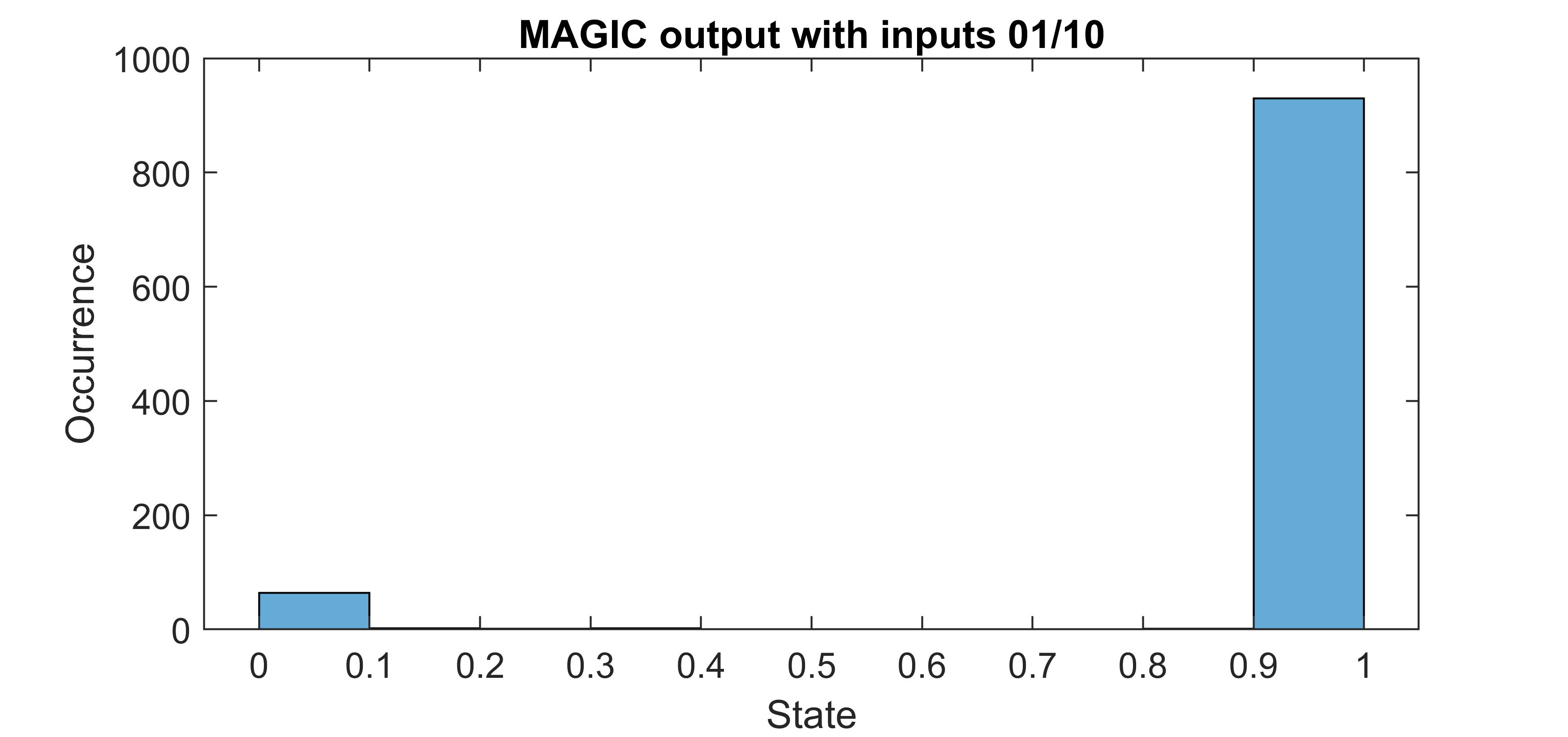}
    \caption[\gls{magic}: State distribution of output memristor]{\gls{magic}: State distribution of output memristor after logic operation for inputs `01' and `10'. }
    \label{fig:MAGIC_01_vari}
\end{figure}

Although the \gls{magic} gate was designed according to the constraints given in the original publication, i.e., Equation~(1)~in~\cite{Kvatinsky2014magic}, the resulting NOR-gate does not work for the inputs `01' and `10', which should yield `0'. This is due to the fact, that the second constraint on applied voltage, i.e., Equation~(2)~in~\cite{Kvatinsky2014magic} cannot be fulfilled at the same time as Equation~(1) with the parameter set measured for the memristors used in this work.
This is manifested in a very low probability of $6.4$\,\% correct output generation in cases `01' and `10', as {it is} shown in \Cref{fig:MAGIC_01_vari}.
Simulations of cases `00' and `11' yielded probabilities of $95.2$\,\% and $87.2$\,\%, respectively.

\gls{magic} is not prone to distortion of outputs by leakage. This is caused by the inherent function of \gls{magic} which is a selective {RESET} of the output memristor, which is initialized to \gls{lrs} prior to logic operation. Since no leakage was observed after a RESET process, \gls{magic} NOR-gates are not influenced by this effect.

\subsection{FELIX}
\label{sec:case_FELIX}
{Fast and Energy-efficient Logic in memory}~(\gls{felix}) can be used to implement different logic functions using the same structure, such as {NOR}, {NAND} or the logical minority function, as {it is} shown in \cite{Gupta2018}. These functions are constructed very similar to \gls{magic} gates. 
However, \gls{felix} can also be used to implement an {OR} function. 
\begin{2Detailed}
\red{
As {it} can be seen in \Cref{fig:FELIX_circ}, a \gls{felix} OR-gate has 
{a similar}
structure
as a \gls{magic}-gate, but the ground and $V_\mathrm{0}$ connections are reversed and the output memristor is initialized to \gls{hrs} prior to logic operation.}
\end{2Detailed}
\Cref{tab:logic_param} gives the logic parameters used for the simulation of \gls{felix} gates.

\begin{figure}[b!]
    \centering
    \includegraphics[width=0.8\linewidth]{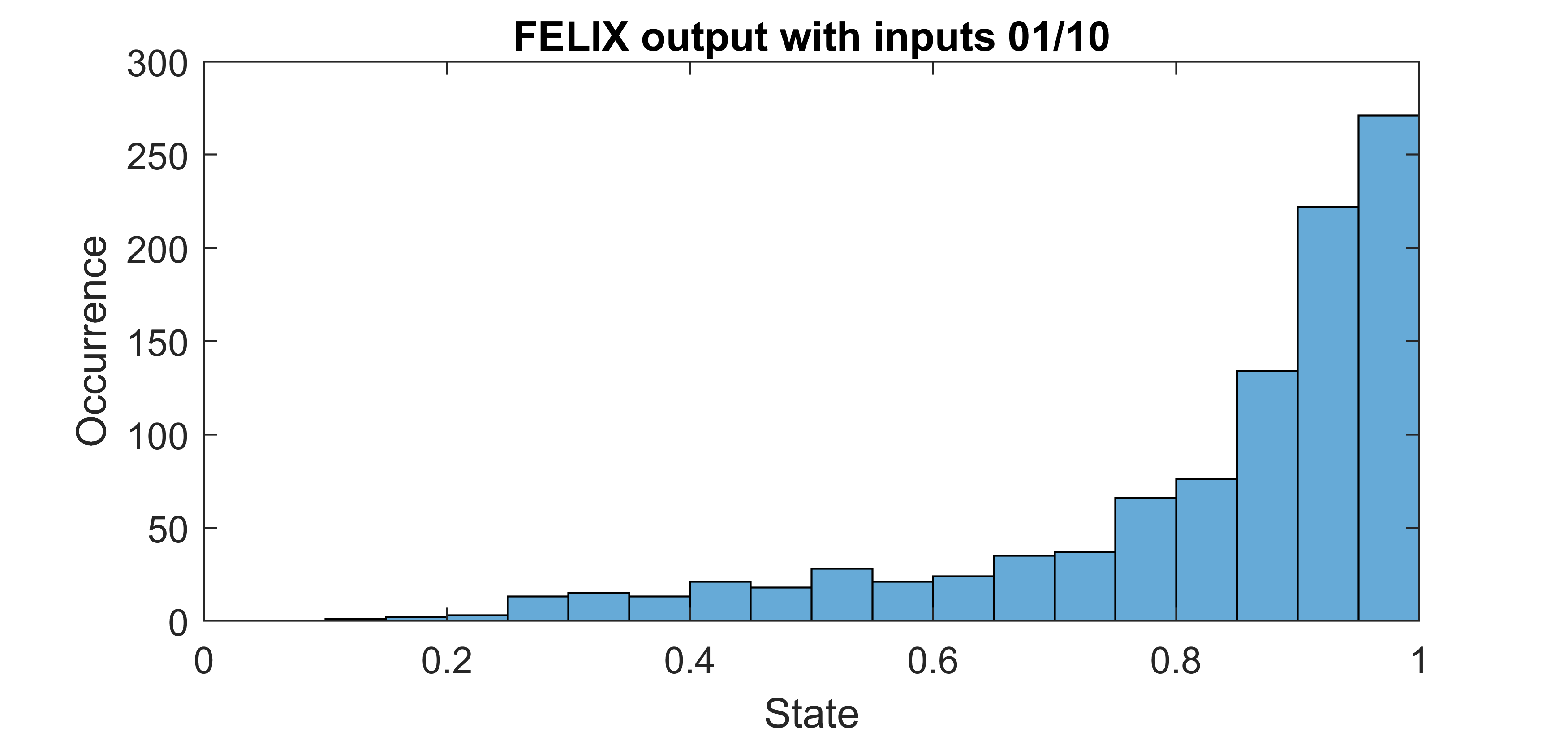}
    \caption[\gls{felix}: State distribution of output memristor for inputs ‘01’ and ‘10’]{\gls{felix}: State distribution of output memristor after logic operation for inputs ‘01’ and ‘10’.}
    \label{fig:FELIX_vari_01}
\end{figure}

\begin{figure}[b!]
    \centering
    \includegraphics[width=0.8\linewidth]{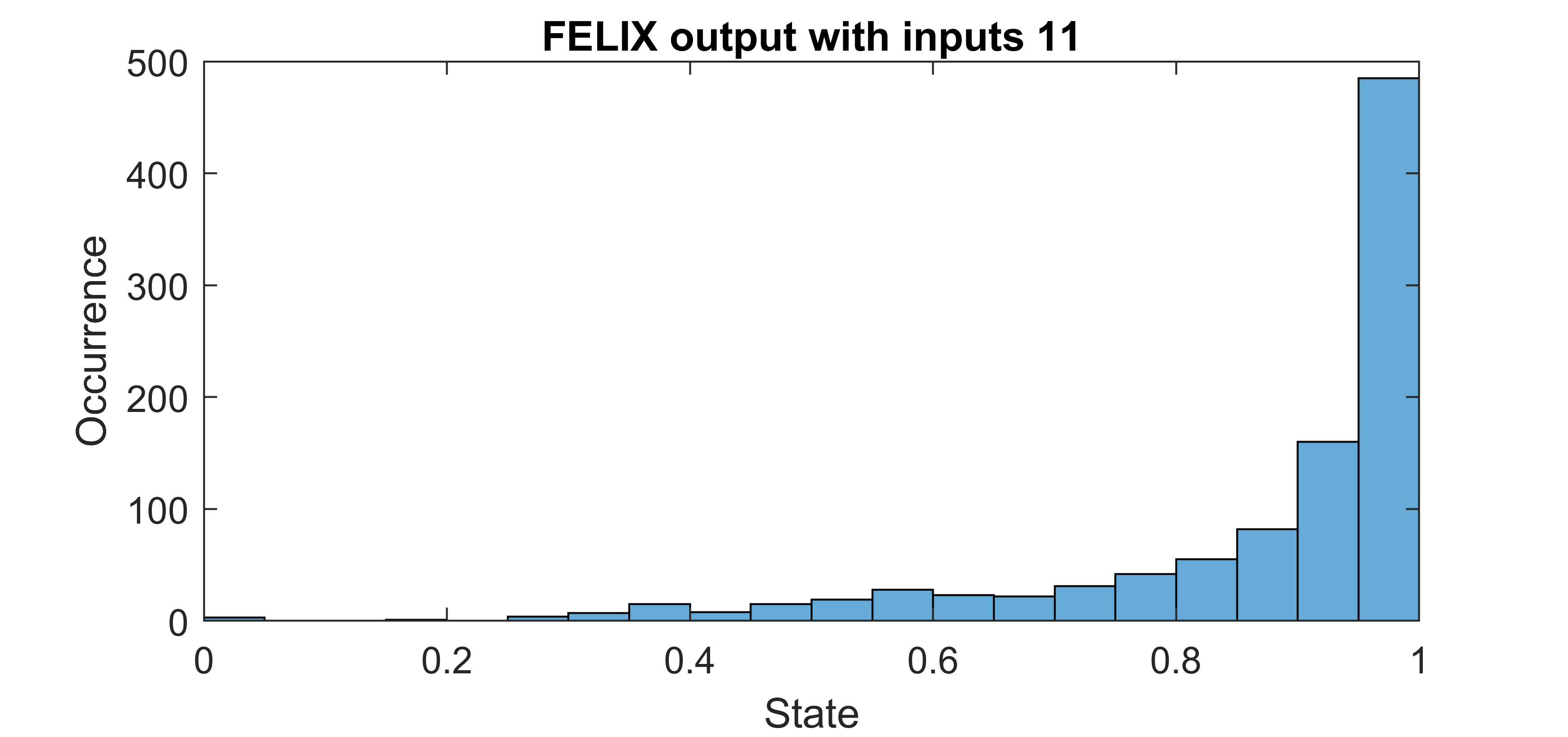}
    \caption[\gls{felix}: State distribution of output memristor for input ‘11’]{\gls{felix}: State distribution of output memristor after logic operation for input ‘11’.}
    \label{fig:FELIX_vari_11}
\end{figure}

\Cref{fig:FELIX_vari_01} shows the histogram of the simulated output states for the inputs ‘01’ and ‘10’. The two cases do not need to be distinguished, since \gls{felix} is symmetrical regarding the inputs. The gate gave correct outputs in $91.4$\,\% of the simulated cases. \Cref{fig:FELIX_vari_11} shows the histogram of the simulated output states for the input ‘11’. The gate gave correct outputs in $94.8$\,\% of the simulated cases. With $100$\,\% correctness for the input ‘00’ this calculates to an overall correctness of $94.4$\,\%.

\begin{figure}[t!]
    \centering
    \includegraphics[width=\linewidth]{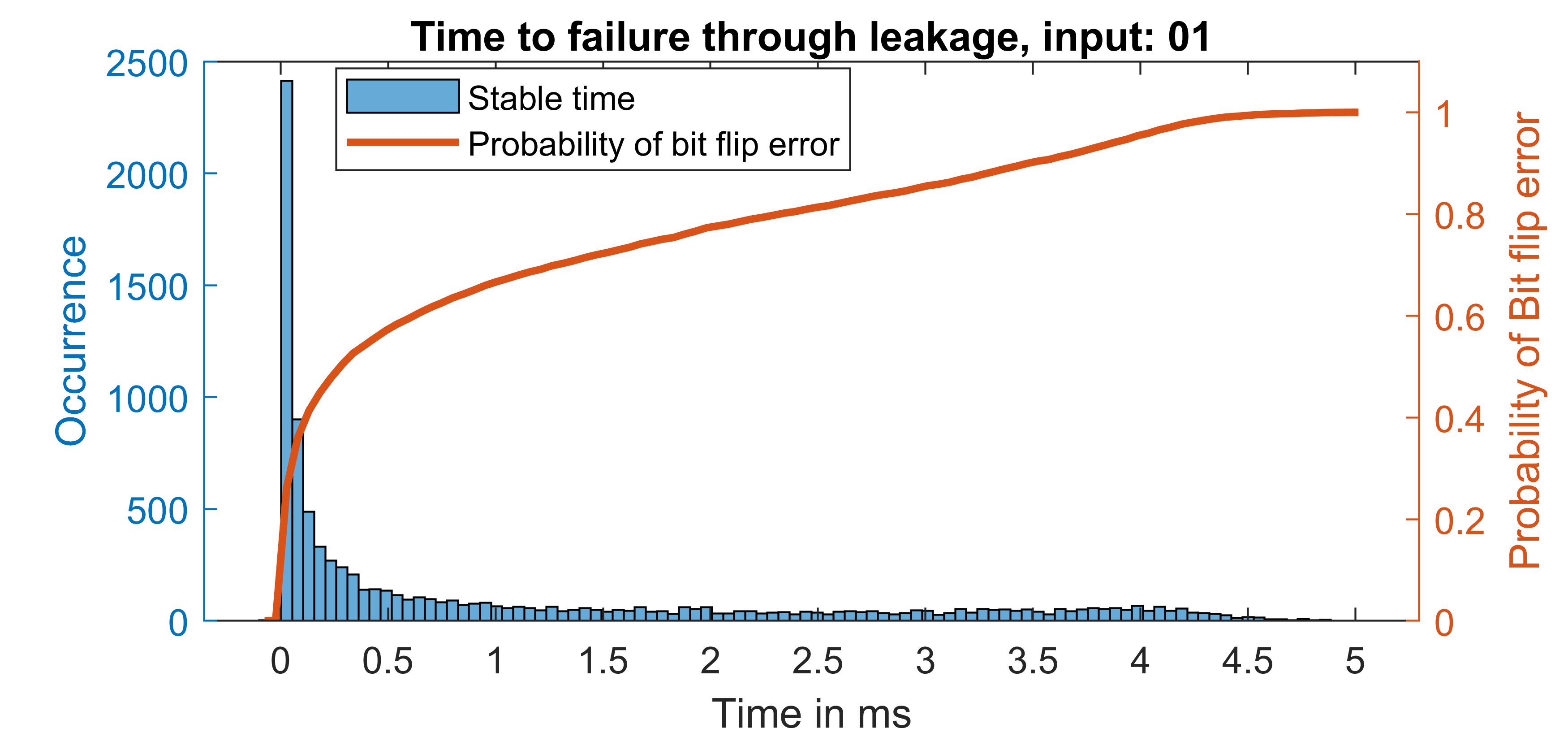}
    \caption[\gls{felix}: Distribution of stable output time for input `01']{\gls{felix}: Distribution of stable output time for input `01'. The orange line gives the accumulated portion of incorrect outputs over time.}
    \label{fig:FELIX_leak_01}
\end{figure}

\begin{figure}[t!]
    \centering
    \includegraphics[width=\linewidth]{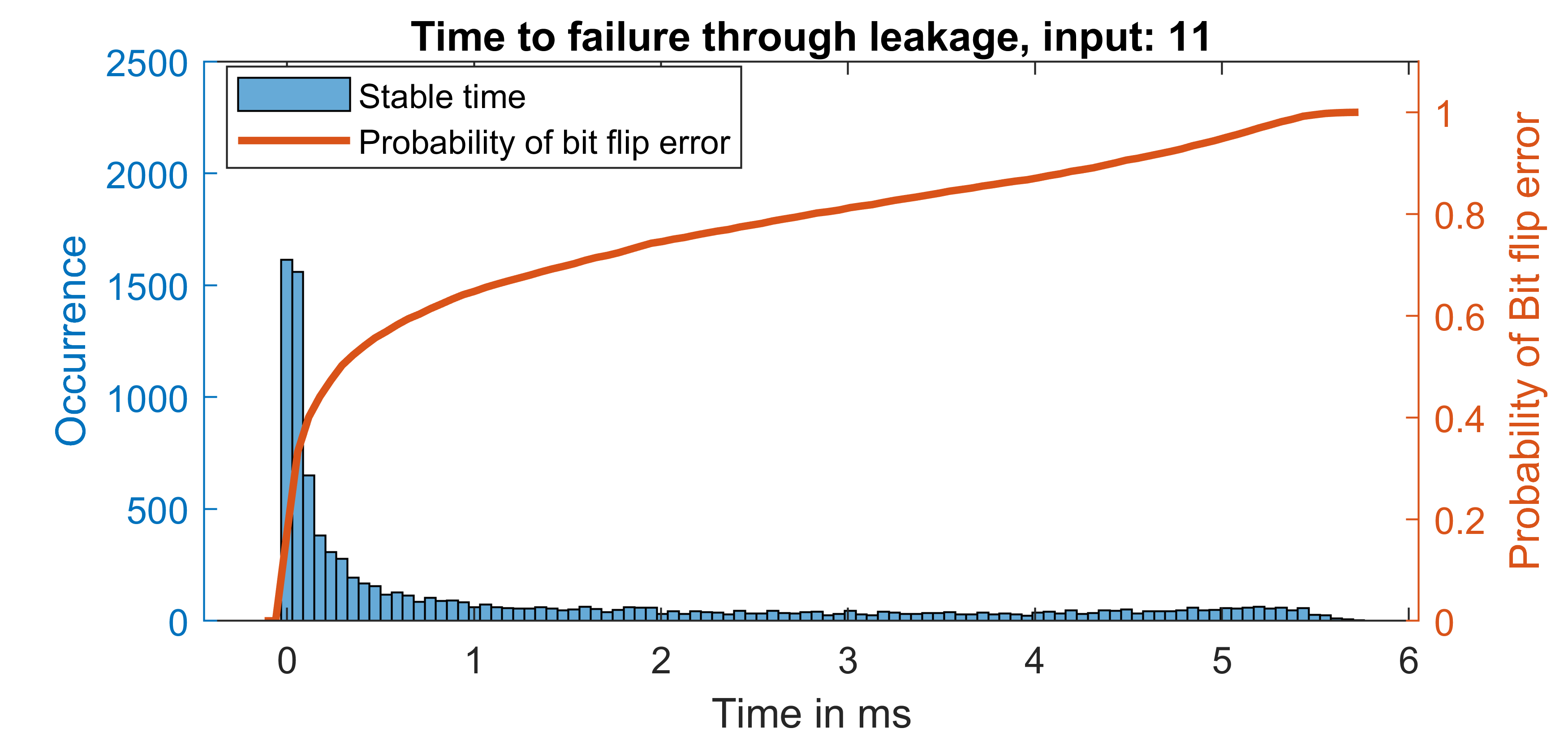}
    \caption[\gls{felix}: Distribution of stable output time for input `11']{\gls{felix}: Distribution of stable output time for input `11'. The orange line gives the accumulated portion of incorrect outputs over time.}
    \label{fig:FELIX_leak_11}
\end{figure}

Leakage affects the output of a \gls{felix} NOR-gate for the inputs ‘01’, ‘10’ and ‘11’, where the cases ‘01’ and ‘10’ are again symmetrical and thus indistinguishable in simulation.
Figures~\ref{fig:FELIX_leak_01} and~\ref{fig:FELIX_leak_11} give the histograms of the time the output was stable during the simulations for the inputs ‘01’,‘10’ and ‘11’, respectively. The orange line in the figures again gives the accumulated relative portion of all outputs being distorted by leakage.
For the input ‘01’ and ‘10’,  $>90$\,\% of the calculated outputs remain correct until $14.1$\,$\mu$s after the operation, for $>99$\,\% the stable time is $3.92$\,$\mu$s.
With the input ‘11’ for $>90$\,\% the stable time is $12.3$\,$\mu$s, for $>99$\,\% it is $0.50$\,$\mu$s.

\subsection{TMSL}
\label{sec:case_TMSL}
The {Three Memristor Stateful Logic}~(\gls{tmsl}) \cite{huang2016reconfigurable} uses a work resistor $R_\mathrm{G}$, as in \gls{imply}, but uses a separate output memristor, as in \gls{magic} and \gls{felix}. 
\Cref{tab:logic_param} gives the logic parameters for the simulation of \gls{tmsl} gates.

\begin{figure}[t!]
    \centering

\def\xlen{4mm}


\begin{circuitikz}[scale=.68,transform shape]
	
	\draw (0,0)node[below]{\large $V_\mathrm{cond}$} to[short,o-](0,1) to [Mr,l_=$in_\mathrm{1}$] (0,3)
	(0,1) to[short,*-](2,1) to[Mr,l_=$in_\mathrm{2}$] (2,3)
	(0,3) to[short,-*](2,3) to[short,-*](4,3)
	(4,1) node[rground]{} to[Mr,l_=$out$] (4,3)
	(4,3) to[short,-](5,3) to[R,l_=$R_\mathrm{G}$] (7,3) 
	(7,3) to[short,-o](7.5,3) node[above]{\large $V_\mathrm{set}$};
\end{circuitikz}

    \caption[Circuit of 2-bit \gls{tmsl} NAND-gate]{Circuit of 2-bit \gls{tmsl} NOR-gate. $in_\mathrm{1}$ and $in_\mathrm{2}$ hold the inputs, $out$ is initialized to \gls{hrs} and holds the output after logic operation.}
    \label{fig:TMSL_circ}
\end{figure}
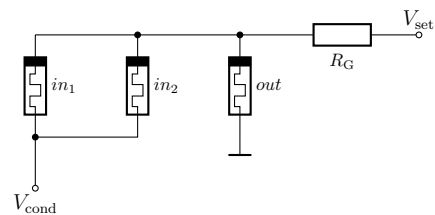

\begin{table}[b!]
    \centering
    \caption[Logic design parameters]{Design parameters for the investigated logics. $T$ denotes the execution time of one logic operation.}
    \begin{tabular}{|cc|c|}
        \hline
        Logic & Parameter & Value \\
        \hline
        \gls{imply} & $T$ & $50$\,$\mu$s \\
                    & $V_\mathrm{set}$ & $0.6$\,V\\
                    & $V_\mathrm{cond}$ & $0.4$\,V\\
                    & $R_\mathrm{G}$ & $40$\,k$\Omega$\\
        \hline
        \gls{magic} & $T$ & $10$\,ms \\
         \& \gls{felix}            & $V_\mathrm{0}$ & $1$\,V \\
        \hline
        \gls{tmsl}  & $T$ & $100$\,$\mu$s \\
                    & $V_\mathrm{set}$ & $1$\,V \\
                    & $V_\mathrm{cond}$ & $0.5$\,V \\
                    & $R_\mathrm{G}$ & $40$\,k$\Omega$ \\
        \hline
    \end{tabular}
    \label{tab:logic_param}
\end{table}

\begin{figure}[t!]
    \centering
    \includegraphics[width=0.8\linewidth]{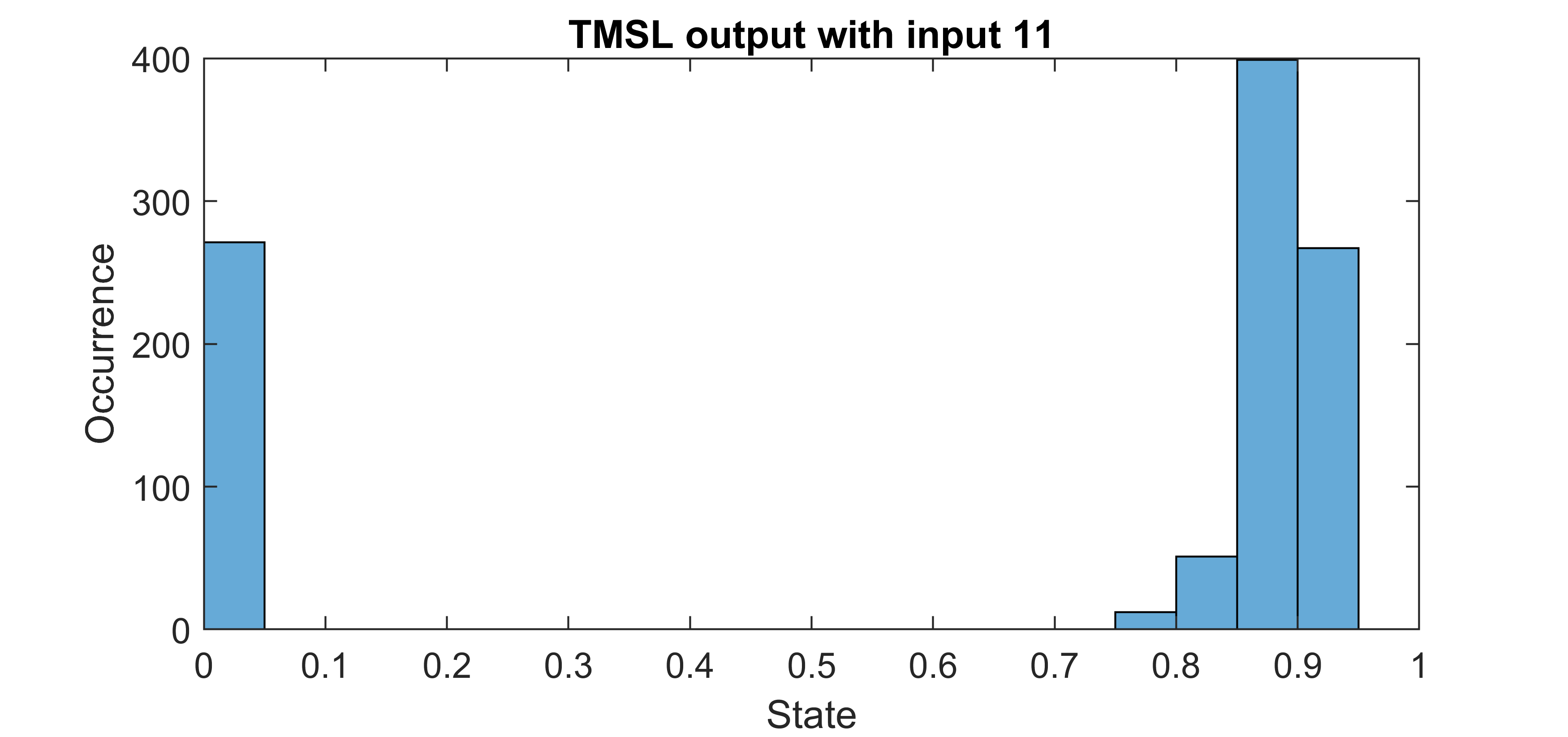}
    \caption[\gls{tmsl}: State distribution of output memristor after logic operation for input ‘11’]{\gls{tmsl}: State distribution of output memristor after logic operation for logical inputs ‘11’, i.e. both input memristors are in \gls{hrs}. Note, that the NAND operation should yield ‘0’ in this case, i.e., memristor state 1.}
    \label{fig:TMSL_vari_11}
\end{figure}

\begin{figure}[t!]
    \centering
    \includegraphics[width=\linewidth]{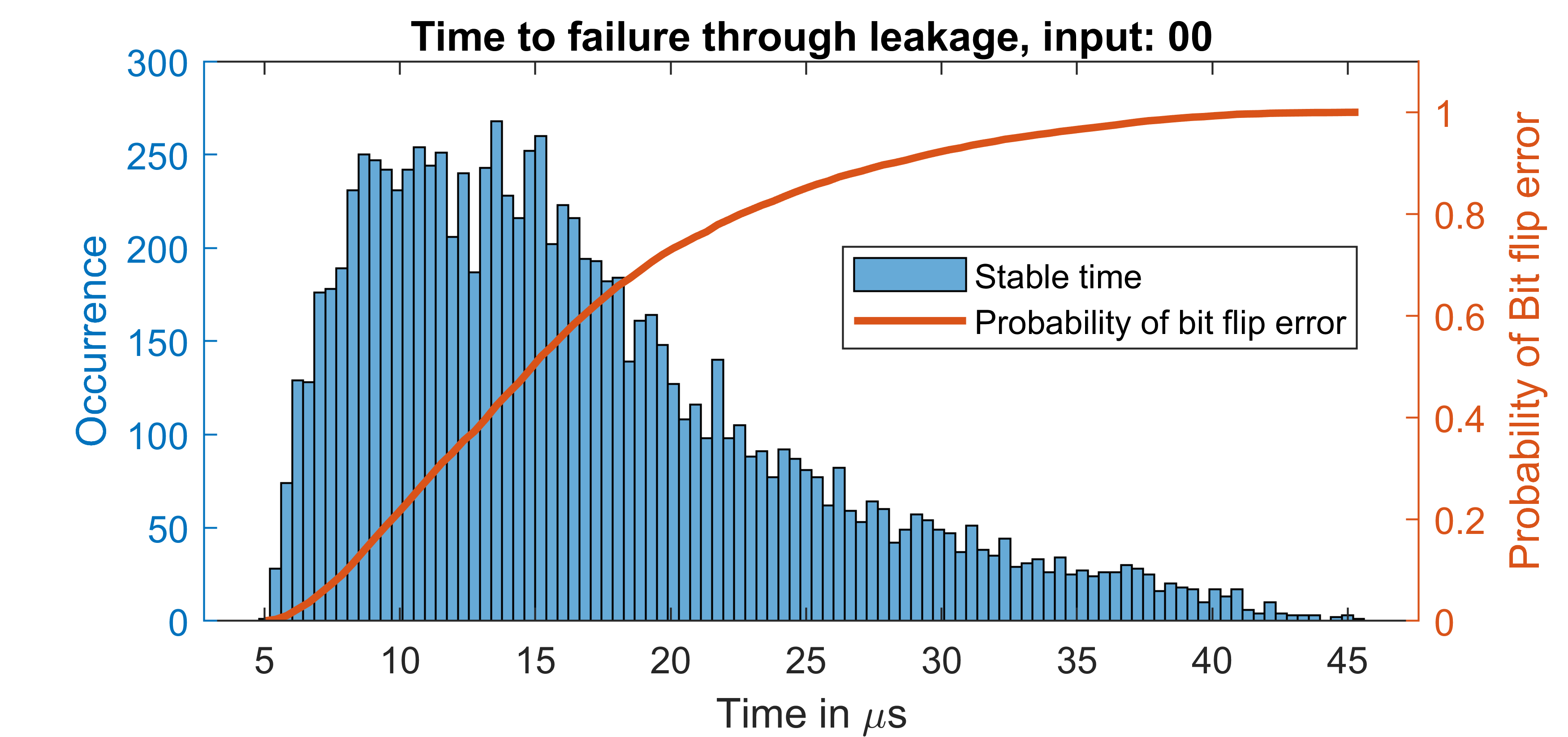}
    \caption[\gls{tmsl}: Distribution of stable output time for input `00']{\gls{tmsl}: Distribution of stable output time for input `00'. The orange line gives the accumulated portion of incorrect outputs over time.}
    \label{fig:TMSL_leak_00}
\end{figure}

\Cref{fig:TMSL_vari_11} shows the histogram of output memristor states observed during simulations for the input ‘11’. As {it} can be seen, $27.1$\,\% of the results are incorrect. With $100$\,\% correct results for the inputs ‘00’, ‘01’ and ‘10’, this calculates to an overall correctness of $93.2$\,\%. Since in \gls{tmsl} the output memristor is initialized to \gls{hrs} and only SET if the input is ‘00’, this denotes the only case that is influenced by leakage. \Cref{fig:TMSL_leak_00} gives the histogram of the stable output time for this input. For $>90$\,\% correct outputs the maximum stable time is $8.21$\,$\mu$s, for $>99$\,\% it is $5.95$\,$\mu$s.

\subsection{Summary}
\label{sec:case_summary}
The presented case studies show, that all of the four most popular stateful memristive logics, i.e., \gls{imply}, \gls{magic}, \gls{felix} and \gls{tmsl},
are negatively influenced by parameter variations and the leakage effect. Even though the gates have been designed according to the given design rules. 
\Cref{tab:logic_performance} gives a summary of the observed performance in terms of output probabilities, \Cref{tab:leak_performance} shows the $90$\,\%, $99$\,\%, average and median output stable times for all investigated logics.
{Note that in \Cref{tab:leak_performance},
if more than one input case was influenced by leakage, the minimum times are given. $t_\mathrm{avg}$ and $t_\mathrm{med}$ denote the average and median stable time of all inputs affected by leakage, respectively. \gls{magic} is not listed in the table since in our technology it is structurally not prone to leakage, as explained in \Cref{sec:case_MAGIC}.
}

\begin{table}
    \centering
    \caption[Probability of correct output for the investigated logics]{Probability of correct output for each of the input combinations (‘00’, ‘01’, ‘10’ and ‘11’) for the investigated logics. }
    \begin{tabular}{|c|c|c|c|c|c|}
    \hline
        Logic & ‘00’ & ‘01’ & ‘10’ & ‘11’ & Overall\\
    \hline
        \gls{imply} & $86.4$\,\% & $100$\,\%  & $100$\,\%  & $100$\,\%  & $96.6$\,\% \\
        \gls{magic} & $95.2$\,\% & $6.4$\,\%  & $6.4$\,\%  & $87.2$\,\% & $48.8$\,\%\\
        \gls{felix} & $100$\,\%  & $91.4$\,\% & $91.4$\,\% & $94.8$\,\% & $94.4$\,\% \\
        \gls{tmsl}  & $100$\,\%  & $100$\,\%  & $100$\,\%  & $72.9$\,\% & $93.2$\,\% \\
    \hline
    \end{tabular}
    \label{tab:logic_performance}
\end{table}

\begin{table}
    \centering
    \caption[Stable output times of the investigated logics]{Stable output times of the investigated logics. $t_\mathrm{90}$ and $t_\mathrm{99}$ denote the time duration for validity of respectively $90$\,\% and $99$\,\% of the outputs. 
    Only initially correct outputs have been considered. 
    }
    \begin{tabular}{|c|c|c|c|c|}
        \hline
        Logic & $t_\mathrm{90}$ in $\mu$s & $t_\mathrm{99}$ in $\mu$s & $t_\mathrm{avg}$ in $\mu$s & $t_\mathrm{med}$ in $\mu$s\\
        \hline
        \gls{imply}  & $16.9$ & $1.98$ & $25.04$ & $27.08$ \\
        \gls{felix}  & $12.3$ & $0.5$ & $1170.5$ & $311.5$ \\
        \gls{tmsl}   & $8.2$ & $5.95$ & $16.8$ & $15.1$ \\
        \hline
    \end{tabular}
    \label{tab:leak_performance}
\end{table}

\section{Conclusion \& Outlook}
\label{chap:conclusion}

The work at hand presents an enhanced model for resistance switching type memristors or {ReRAMs}, {which was fitted to {KNOWM} \gls{sdc} memristors.} 
For the first time in the field of behavioral {ReRAM} models, the so-called leakage effect, i.e., a drift of the memristor state in absence of any stimulus, is reproduced. 
The leakage effect was observed and modeled to shift 
the resistance of a memristor
back to higher values after a {SET} process, i.e., after the memristor changed from higher resistance to lower resistance. 
{This drift happens in an exponential manner over an average time span ($5\tau$) of $51.5$\,s.}
Furthermore, the model supports simulation of parameter variations, which can be used to model both cycle-to-cycle as well as device-to-device variation.
These variations of model parameters were observed to account for up to $23.3$\,\% {of the nominal value.}
The model is completely implemented in \gls{spice} without the need of additional, e.g., procedural, software or solver engines.
The simulation inaccuracy of the leakage effect is 1.1\,\% on average and $13.4$\,\% in the worst case over 20 tested cases. The simulation inaccuracy of resistance dynamics is 4.6\,\% on average and $8.2$\,\% in the worst case over 100 tested cases. 
{By incorporating the leakage effect and the statistical variation of model parameters, this model enables more significant, robust and reliable simulation of circuits and systems containing {ReRAM} devices.}

Based on the presented model and fitting, a set of case studies was conducted, which revealed a negative impact of parameter variations and the leakage effect on four of the most popular stateful memristive logics, namely \gls{imply}, \gls{magic}, \gls{felix} and \gls{tmsl}. The work at hand provides a guideline on how to assess logic robustness and state stability times using the presented model. The probability of correct output generation was simulated to be as low as $48.8$\,\% and the leakage effect was simulated to reduce the stable time of the output memristor state to as low as $0.5$\,$\mu$s, if stable outputs with a probability $\geq 99$\,\%  are desired.

{This work shows, that the evaluation of memristive circuits and systems based on nominal, i.e., static parameter sets for memristor models is insufficient. Parameter variations and the leakage effect need to be taken into account in order to assess the robustness, and in extreme cases even the function, of memristive circuits and systems. These effects proved to degrade the performance of the most popular memristive logics considerably and therefore cannot be neglected.}

The presented model gives a versatile base to conduct further research in stochastic behavior of resistance switching devices and memristors in general. 
For example, the statistical properties of the leakage effect could be investigated and a thorough statistical analysis of memristive behavior around threshold voltages 
and currents could be conducted. 
{Large scale statistical analysis could be performed to refine the modeling of parameters with uniform distribution and work-out more sophisticated distribution models.}
{Future work could also investigate
if there is any influence due to parameter variation of resistance dynamics on the variation of the leakage effect.}
To further increase the statistical relevance, large scale stochastic analysis of all discussed effects could be conducted.
Such research requires elaborate and sophisticated designs regarding experimentation environments and measurement methods for which this work can act as a base in terms of instrumentation and methodology.
Measured data availability in the field of memristive circuits and systems is very scarce in general and ongoing statistical examination of memristor behavior is necessary. This work strives to take one step forward in that direction.
\section*{Model Source File}
This model will be open-sourced and available to the public upon the acceptance of the paper.
\bibliographystyle{IEEEtran}
\bibliography{references,setups/nima,setups/memristor_lit}
\vspace{-3.7em}
\begin{IEEEbiography}
[{\includegraphics[width=1in,height=1.13in,trim =1mm 3mm 1mm 13mm, clip,keepaspectratio]{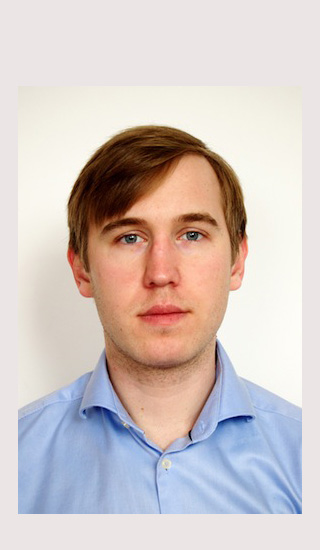}}]{David Radakovits} 
received his M.Sc. (2020) in Embedded Systems and B.Sc. (2017) in Electrical Engineering and Information Technology from Technische Universit\"at Wien, Vienna University of Technology. His research interest includes the areas of memristor-based circuits and systems, hardware security, machine learning and embedded systems in which he published several articles in {IEEE} journals and conferences.
\end{IEEEbiography}
\vspace{-3.7em}
\begin{IEEEbiography}
[{\includegraphics[width=1in,height=1.25in,clip,keepaspectratio]{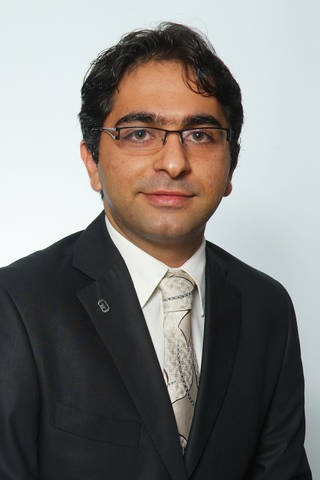}}]{Nima TaheriNejad} 
(S'08-M'15) received his Ph.D. degree in electrical and computer engineering from The University of British Columbia (UBC), Vancouver, Canada, in 2015.
He is currently a ``Universit\"{a}tsassistant'' at the TU Wien (formerly known as Vienna University of Technology), Vienna, Austria, where his areas of work include in-memory computing and self-awareness in resource-constrained cyber-physical and embedded systems, and health-care. He has published two books and more than 60 peer-reviewed articles. Dr. Taherinejad has also served as a reviewer, an editor, an organizer, and the chair for various journals, conferences, and workshops.
He has received several awards and scholarships from universities
and conferences he has attended.
\end{IEEEbiography}


\end{document}